\documentclass[11pt]{amsart}

\usepackage[foot]{amsaddr}
\usepackage[T1]{fontenc}       
\usepackage[textsize=tiny]{todonotes}
\usepackage{upgreek}           
\usepackage[colorlinks=true, citecolor=blue, linkcolor=blue]{hyperref}
\usepackage{caption}
\usepackage{graphicx}
\usepackage[marginratio=1:1,height=584pt,width=360pt,tmargin=117pt]{geometry}
\usepackage[sort&compress,numbers,super]{natbib}

\usepackage[right, modulo]{lineno}
\linenumbers

\newcommand{\beginsupplement}{%
  \setcounter{figure}{0}%
  \renewcommand{\figurename}{\textbf{Extended Data Figure}}%
  \newgeometry{width=7.5in, height=9in}%
}

\renewcommand{\figurename}{\textbf{Figure}}

\author{Rees F. Garmann$^{1,\dag}$, Aaron M. Goldfain$^{1,\dag}$, and Vinothan N. Manoharan$^{1,2}$}
\address{$^1$Harvard John A. Paulson School of Engineering and Applied
  Sciences, Cambridge, MA 02138 USA}
\address{$^2$Department of Physics, Harvard University, Cambridge, MA
  02138 USA}
\address{$^\dag$equal contribution}
\email{vnm@seas.harvard.edu}

\title[Self-assembly kinetics of individual viral
capsids]{Measurements of the self-assembly kinetics of individual viral
  capsids around their RNA genome}

\begin{document}
\maketitle


\textbf{The formation of a viral capsid---the highly-ordered protein
  shell that surrounds the genome of a virus---is the canonical example
  of self-assembly~\cite{caspar_physical_1962}. The capsids of many
  positive-sense RNA viruses spontaneously assemble from \textit{in
    vitro} mixtures of the coat protein and
  RNA~\cite{bancroft_formation_1967}. The high yield of proper capsids
  that assemble is remarkable, given their structural complexity: 180
  identical proteins must arrange into three distinct local
  configurations to form an icosahedral capsid with a triangulation
  number of 3 ($T=3$)~\cite{caspar_physical_1962}. Despite a wealth of
  data from structural studies~\cite{harrison_tomato_1978,
    fisher_ordered_1993, dai_situ_2017} and
  simulations~\cite{berger_local_1994, schwartz_local_1998,
    elrad_encapsulation_2010, perlmutter_pathways_2014,
    dykeman_solving_2014}, even the most fundamental questions about how
  these structures assemble remain unresolved. Experiments have not
  determined whether the assembly pathway involves aggregation or
  nucleation, or how the RNA controls the process. Here we use
  interferometric scattering
  microscopy~\cite{jacobsen_interferometric_2006,
    young_2018_quantitative} to directly observe the \textit{in vitro}
  assembly kinetics of individual, unlabeled capsids of bacteriophage
  MS2. By measuring how many coat proteins bind to each of many
  individual MS2 RNA strands on time scales from 1~ms to 900~s, we
  find that the start of assembly is broadly distributed in time and is
  followed by a rapid increase in the number of bound proteins. These
  measurements provide strong evidence for a nucleation-and-growth
  pathway. We also find that malformed structures assemble when multiple
  nuclei appear on the same RNA before the first nucleus has finished
  growing. Our measurements reveal the complex assembly pathways for
  viral capsids around RNA in quantitative detail, including the
  nucleation threshold, nucleation time, growth time, and constraints on
  the critical nucleus size. These results may inform strategies for
  engineering synthetic capsids~\cite{butterfield_evolution_2017} or for
  derailing the assembly of pathogenic
  viruses~\cite{deres_inhibition_2003}.}


\begin{figure}
	\centering \includegraphics{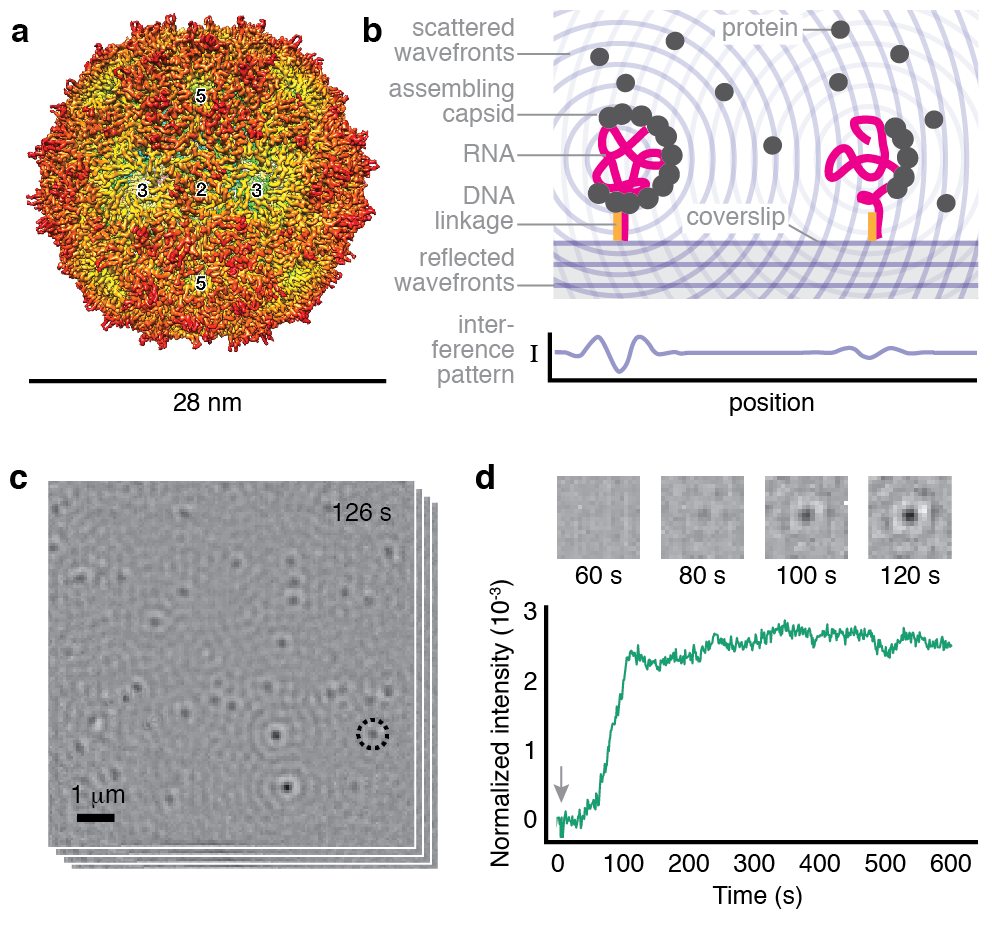}
	\caption{\textbf{Overview of the system and measurement.} (a) A
    structural model of the MS2 capsid from cryo-electron microscopy
    data (reproduced from Ref.~\citenum{dai_situ_2017}) reveals its
    icosahedral structure with 2-, 3-, and 5-fold symmetry axes. (b) We
    inject a solution of MS2 coat-protein dimers over a coverslip on
    which MS2 RNA strands are tethered by DNA
    linkages~\cite{garmann_simple_2015}. As the proteins bind to the
    RNA, the resulting particles scatter light. Owing to destructive
    interference between the scattered light and a reference beam, the
    particles appear as dark, diffraction-limited spots. (c) We monitor
    many individual assembling particles in parallel. A typical image,
    taken 126~s after adding 2 $\upmu$M dimers and representing an
    average of 1,000 frames taken at 1,000 frames/s, shows multiple
    spots. (d) The intensity of a spot as a function of time reveals the
    assembly kinetics of an individual particle. Top: time-series of
    images for the circled spot in (c). Bottom: kinetic trace for the
    same spot using a 1,000-frame average of data taken at 1,000
    frames/s. The arrow indicates when we inject the coat protein.}
    	\label{fig:overview} 
\end{figure}

We work with MS2 (Fig.~\ref{fig:overview}a) because it is a non-trivial
model system for understanding capsid assembly: Complete 28-nm capsids
can be assembled \textit{in vitro} from the coat proteins and
RNA~\cite{sugiyama_ribonucleoprotein_1967}; the assembled capsids have
$T=3$ (90 coat-protein dimers)~\cite{toropova_three_2008}, such that
they must compete with many possible malformed structures; and the RNA
is suspected to play an important role in the assembly
process~\cite{hohn_role_1969, romaniuk_1987, beckett_roles_1988,
  stockley_simple_2007, borodavka_evidence_2012}. Experiments probing
assembly in bulk solution have shown that specific RNA sequences might
initiate assembly by binding the first few coat
proteins~\cite{borodavka_evidence_2012}. But because such measurements
probe an ensemble of particles in possibly different stages of assembly,
they can obscure important features of the assembly pathway. The equally
important question of how that pathway can be derailed---leading to the
often-overlooked minority of malformed structures observed in bulk
assembly of bacteriophages~\cite{hohn_role_1969} and other
viruses~\cite{sorger_structure_1986, kler_scaffold_2013,
  garmann_assembly_2014}---also remains unresolved.

Our interferometric scattering experiments address these questions
because they probe the assembly of individual capsids
(Fig.~\ref{fig:overview}b--d). As described in Methods, each assembling
particle produces a diffraction-limited spot in the field of view.
Because continuous background correction in our measurement renders the
RNA invisible, the final signal depends only on the number of proteins
in the assembling particle. Thus, the time trace of the intensity for
each spot gives a direct measure of the assembly kinetics of an
individual particle.

We must measure the intensity of each spot with both high sensitivity
and high dynamic range, because the capsids scatter weakly, and
estimates of the assembly times range from
seconds~\cite{beckett_roles_1988} to hours~\cite{beckett_roles_1988,
  borodavka_evidence_2012}. Our apparatus (Extended Data
Fig.~\ref{fig:apparatus}) addresses both of these challenges. Because
the scattering is elastic, we can use high illumination intensities with
minimal risk of photodamage, enabling temporal resolutions of 1~ms. To
simultaneously achieve durations of 900~s, we actively stabilize the
microscope in all three dimensions, ensuring that the signal from the
assembling capsid is larger than the noise due to drift. The sensitivity
is then limited by shot noise. With a 1-s moving average, as shown in
Fig.~\ref{fig:overview}d, the peak-to-peak fluctuations from shot noise
correspond to the intensity of six coat-protein dimers.

\begin{figure}
	\centering \includegraphics{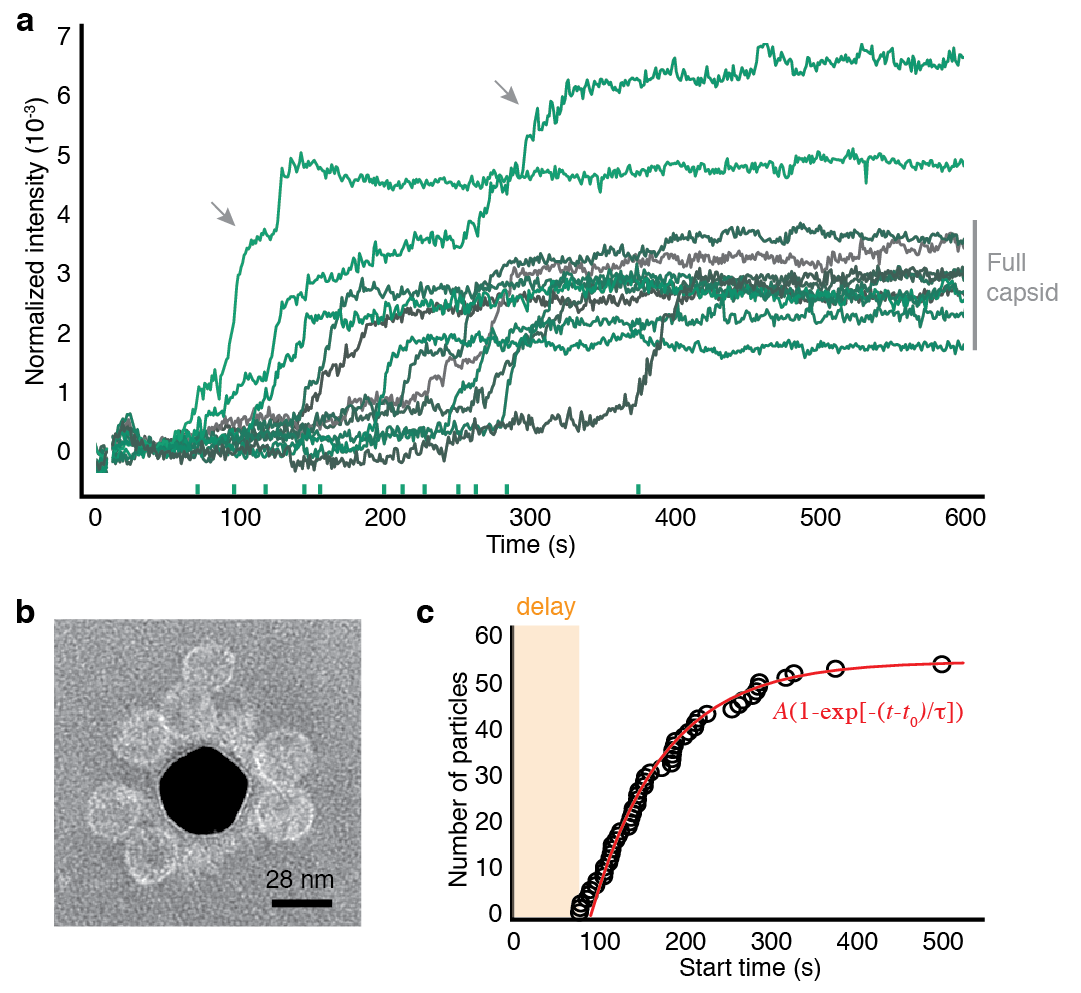}
	\caption{\textbf{Assembly of 2 $\upmu$M coat-protein dimers around
      surface-tethered RNA strands.} (a) Kinetic traces for 12 randomly
    chosen particles. Ticks on the x-axis show the start times. Grey bar
    indicates the intensity range corresponding to wild-type capsids.
    Arrows point to two traces corresponding to overgrown particles. (b)
    A negatively stained TEM image of particles assembled around RNA
    strands tethered to a gold nanoparticle (dark region at center). We
    use a nanoparticle as the substrate because TEM cannot image through
    a coverslip. (c) The cumulative distribution of start times in the
    traces is well fit by an exponential function with $A$ = 56.62 $\pm$
    0.02, $t_0$ = 91.8 $\pm$ 0.2~s, and $\tau$ = 84.3 $\pm$ 0.2~s.
    Uncertainties in the time measurements are smaller than the diameter
    of the circles.}
	\label{fig:MS2assembly}
\end{figure}

In a single experiment, we measure kinetic traces for many assembling
particles in parallel, and we characterize the shape of each trace as
well as variations among traces. When we inject 2 $\upmu$M coat-protein
dimers, we find that most traces have an initial plateau at a low
intensity followed by a rapid rise and a second plateau at higher
intensity (Fig.~\ref{fig:MS2assembly}a and Extended Data
Fig.~\ref{fig:2uMassembly}). A few traces show intermediate plateaus.
Most (40 out of 56) plateau at an intensity consistent with that of a
full, wild-type capsid (Extended Data Fig.~\ref{fig:calibration}), 7 at
a slightly lower intensity, and 9 at a significantly higher intensity.
No such traces are observed when RNA is not tethered to the coverslip
(Supplementary Information). Furthermore, negatively stained
transmission electron microscopy (TEM) images reveal that most of the
structures assembled in control experiments are proper capsids, with a
few partial capsids and larger structures visible
(Fig.~\ref{fig:MS2assembly}b and Extended Data
Fig.~\ref{fig:TEM_tether}). We therefore infer that capsids can indeed
assemble around tethered RNA strands, and that traces that reach
intensities similar to those of wild-type capsids represent the
formation of complete or nearly-complete capsids.

With this understanding, we examine what the traces reveal about the
assembly pathway. A key observation is that assembly is not synchronous:
the `start time', the time at which the intensity rapidly increases,
varies from particle to particle (Fig.~\ref{fig:MS2assembly}a). We find
that the cumulative distribution of start times $t$ is fit well by an
exponential function
$A\left(1-\exp\left[-(t-{t}_{0})/{\tau}\right]\right)$
(Fig.~\ref{fig:MS2assembly}c), where $A$ is the plateau value, $t_0$ is
the delay before the start time of the first particle, and $\tau$ is the
characteristic time (see Methods and Extended Data
Fig.~\ref{fig:nucleation_times}.)

The delay likely results from the combination of diffusion and a
concentration threshold for assembly. We know such a threshold exists
because we see no assembly when we inject 1 $\upmu$M coat-protein dimers
(Extended Data Fig.~\ref{fig:1uMassembly}). Because the threshold is
between 1 and 2 $\upmu$M coat-protein dimers, we expect the delay to be
of the order of the characteristic time for protein to diffuse from the
2 $\upmu$M injected fluid to the surface. Indeed, that time scale is
30--55~s (Supplementary Information), and the observed delay time is
92~s.
 
The distribution of start times, however, does not appear to result from
diffusion. The distribution is broad, with the largest start time
(500~s) an order of magnitude larger than the delay time. The
distribution could result from diffusion-limited growth only if the
protein concentration around each RNA were to vary across the
10-$\upmu$m field of view. But the time for a dimer to diffuse 10
$\upmu$m is only 1~s, much shorter than the median start time.
Furthermore, we estimate that about 1,000 coat-protein dimers are within
1~$\upmu$m of each RNA after the initial delay. At this concentration,
the pool of coat proteins is not significantly depleted by assembly, and
fluctuations in concentration are negligible. We conclude that the
observed kinetic traces do not result from variations in protein
concentration.

Taken together, these findings rule out a diffusion-limited aggregation
pathway and point strongly to nucleation and growth. The exponential
shape of the cumulative distribution of start times suggests a
well-defined free-energy barrier to nucleation with a nucleation time of
$\tau$. Although nucleation models have been used to describe the bulk
assembly kinetics of empty capsids \cite{zandi_classical_2006,
  prevelige_jr_nucleation_1993, zlotnick_theoretical_1999}, and computer
simulations have explored nucleated pathways for capsid assembly around
RNA \cite{elrad_encapsulation_2010, perlmutter_pathways_2014,
  dykeman_solving_2014}, direct experimental evidence for nucleation has
remained elusive. The evidence that we present---the distribution of
start times---cannot easily be extracted from bulk
experiments~\cite{borodavka_evidence_2012}, which average over an
ensemble of particles, or from structural
experiments~\cite{medrano_imaging_2016}, which have coarse temporal
resolution.

Fluctuations in the intensity reveal further information about the
nucleation event. Before the start time, the fluctuations are consistent
with those expected from shot noise, which, as noted above, corresponds
to six dimers at 1-s averaging. This measurement indirectly constrains
the critical nucleus size: we can infer that sub-critical nuclei smaller
than six dimers do not survive for longer than 1~s.

Additional nucleation events may be responsible for assembly going awry
in some of the capsids. Most of the traces with final plateau
intensities higher than that of a full capsid also show intermediate
plateaus at intensities consistent with that of a full capsid
(Fig.~\ref{fig:MS2assembly}a). Such traces suggest that the particle
undergoes a second nucleation event after the first capsid is nearly
complete.

\begin{figure}
	\centering \includegraphics{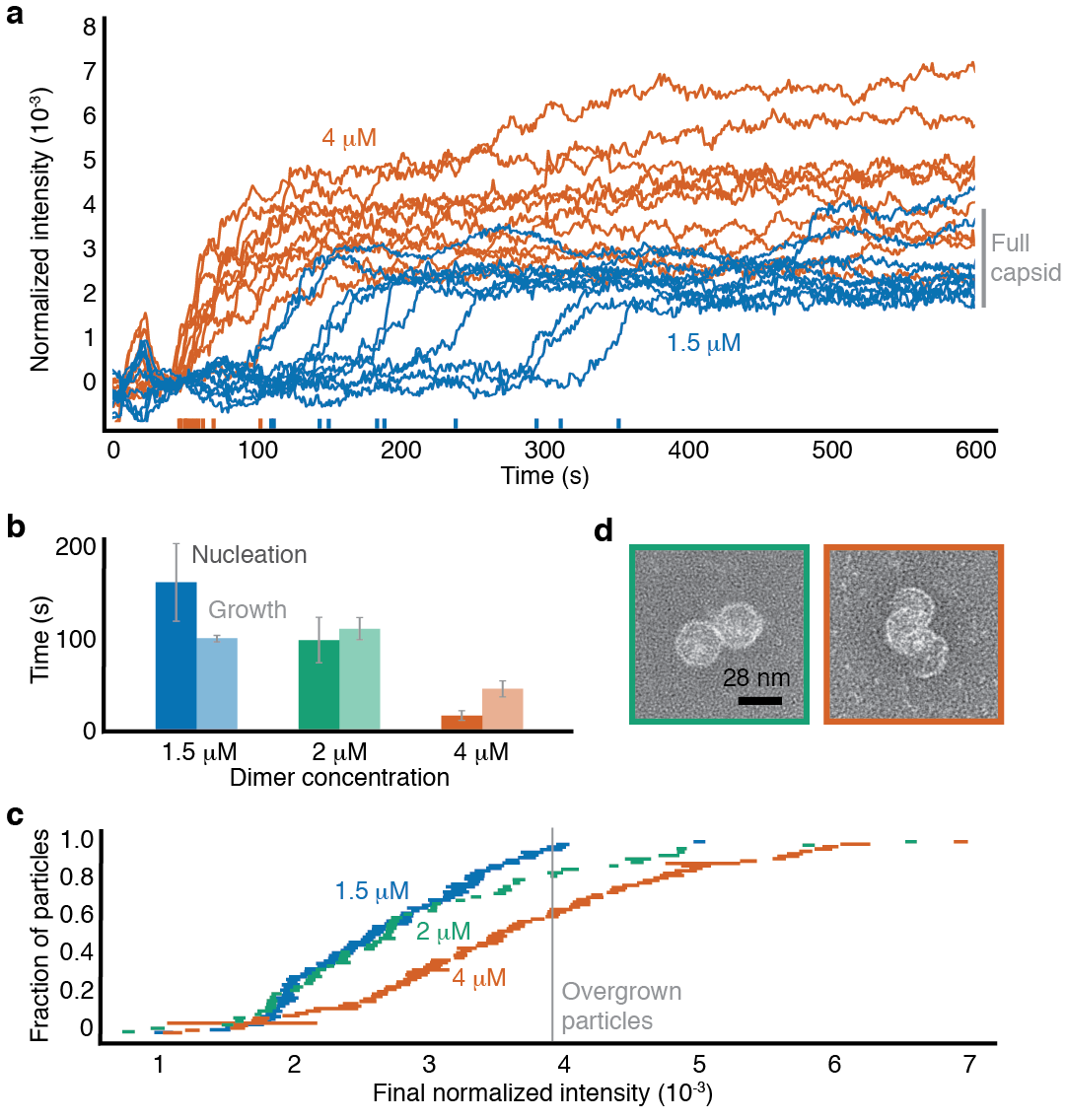}
	\caption{\textbf{Assembly kinetics at different protein
      concentrations.} (a) Kinetic traces for 10 randomly chosen
    particles at 1.5 $\upmu$M and 4 $\upmu$M coat-protein dimers. (b)
    Measured nucleation times and median growth times. Error bars
    represent the standard deviation from three experiments. (c)
    Cumulative distributions of the final intensities show that the
    fraction of overgrown particles increases with protein
    concentration. The length of each horizontal bar is the standard
    deviation calculated from the last 50~s of each trace. (d) TEM
    images of overgrown particles around untethered RNA. Left: an
    attached pair of nearly complete capsids at 2 $\upmu$M protein.
    Right: connected partial capsids at 4 $\upmu$M protein.}
	\label{fig:MS2concentration}
\end{figure}  

To test this hypothesis, we measure the kinetics at different
concentrations of protein (Fig.~\ref{fig:MS2concentration}a, and
Extended Data Figs.~\ref{fig:1.5uMassembly} and \ref{fig:4uMassembly}).
We find that the nucleation time decreases with increasing protein
concentration, from about 160~s at 1.5~$\upmu$M dimers to about 11~s at
4~$\upmu$M (Fig.~\ref{fig:MS2concentration}b). This decrease is
accompanied by an increase in the fraction of overgrown particles, from
5\% at 1.5~$\upmu$M dimers to over 40\% at 4~$\upmu$M
(Fig.~\ref{fig:MS2concentration}c). TEM images of assembly
reactions around untethered RNA (Extended Data Fig.~\ref{fig:TEM_free})
show overgrown particles with sizes corresponding to the final
intensities seen in the kinetic traces. Many of the overgrown particles
consist of bunches of partial or nearly-complete capsids
(Fig.~\ref{fig:MS2concentration}d).

The kinetic traces and images of the overgrown structures suggest a
pathway involving more than one nucleation event. However, many of the
traces at 4 $\upmu$M coat-protein dimers do not show intermediate
plateaus (Fig.~\ref{fig:MS2concentration}a). To understand why, we
measure how long it takes a particle to reach the intensity of a full
capsid after it starts growing (see Methods). We find that these `growth
times' decrease with increasing protein concentration, but less rapidly
than do the nucleation times (Fig.~\ref{fig:MS2concentration}b). When
the nucleation time is smaller than the growth time, as it is in
experiments with 4 $\upmu$M dimers, additional nuclei can form before
the first has time to grow. Under such conditions, most of the kinetic
traces should not---and indeed, do not---show intermediate plateaus.

\begin{figure}
	\centering \includegraphics{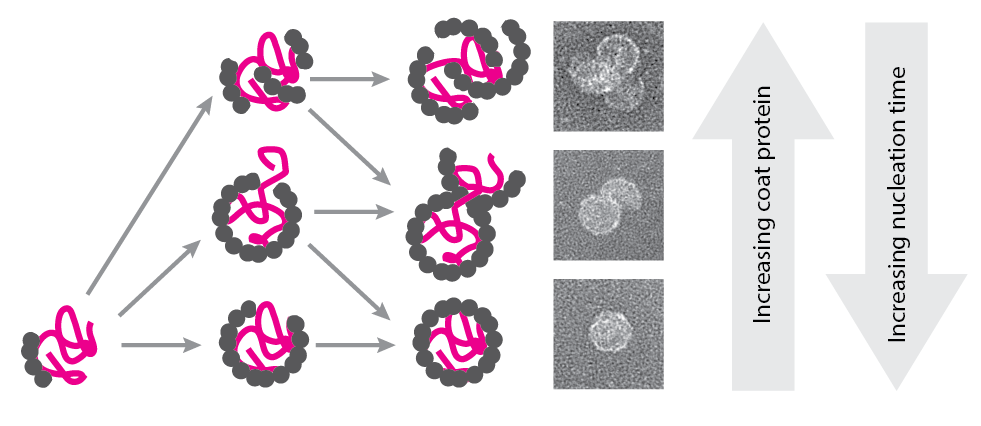}
	\caption{\textbf{Cartoon of the inferred assembly pathways.} First, a
    nucleus of coat proteins forms on the RNA. Bottom row: At low
    protein concentration, no additional nuclei form, and the nucleus
    grows into a proper capsid. Middle row: At higher concentrations, a
    second nucleus can form on an unpackaged part of the RNA, leading to
    a multiplet structure consisting of a nearly-complete capsid
    attached to a second partial or full capsid. Top row: At even higher
    concentrations, multiple nuclei can form and grow, leading to a
    monster structure consisting of many partial capsids. Example TEM
    images of the endpoints of each pathway are shown at right.}
	\label{fig:cartoon}
\end{figure}

Thus, the viral RNA creates a competition between nucleation and growth,
as sketched in Fig.~\ref{fig:cartoon}. Similar scenarios have been
observed in computer simulations of capsid assembly on polymer
scaffolds~\cite{elrad_encapsulation_2010, perlmutter_pathways_2014}, and
may explain the formation of the `monster'~\cite{sorger_structure_1986}
and `multiplet'~\cite{garmann_assembly_2014, kler_scaffold_2013}
structures observed in experiments with other viruses. These structures
are not observed in experiments on the assembly of empty
capsids~\cite{lavelle_phase_2009, pierson_charge_2016}, confirming that
the RNA plays a critical role in the assembly pathways.

Our individual-particle measurements also rule out some competing
pathways. The assembly of proper capsids appears to follow a one-step
nucleation pathway rather than a multi-step
one~\cite{deyoreo_crystal_2013}. Also, the formation of overgrown
structures appears to result from multiple nucleation events rather than
the `spiraling' pathway~\cite{berger_local_1994, schwartz_local_1998}
observed in local rules-based simulations, or the `en-masse'
pathway~\cite{elrad_encapsulation_2010, perlmutter_pathways_2014}
observed in Brownian dynamics simulations. Because our measurements
involve thermally annealed RNA (see Methods), they do not yet resolve
whether the pathway is fine-tuned by local folding patterns in the viral
RNA~\cite{prevelige_follow_2016, song_limits_2017}.

While our observations are specific to \textit{in vitro} assembly, the
observed threshold concentration for nucleation supplies a hypothesis
for viral replication \textit{in vivo}. Because viral RNA that is inside
a capsid cannot be replicated or translated, a virus such as MS2 must
delay encapsidation until its components have been produced in
sufficient quantities. With a threshold for nucleation, assembly (and
encapsidation) would take place only after there is enough viral RNA and
coat protein to form many new virus particles.

Although we expect the assembly pathways to differ for different viruses
and buffer conditions, our measurements of the nucleation time,
threshold, growth time, and subcritical fluctuations in MS2 provide
important constraints on models of assembly. As a result, the structures
of the assembly intermediates and the critical nucleus, which have long
eluded direct imaging methods, might now be inferred through
quantitative comparisons of simulated~\cite{elrad_encapsulation_2010,
  perlmutter_pathways_2014} and measured individual-particle kinetics.
This approach might identify conditions for assembly of synthetic
viruses\cite{butterfield_evolution_2017} and new targets for antiviral
therapies that work by disrupting capsid
assembly~\cite{deres_inhibition_2003}.

\clearpage


\section*{Methods}

\subsection*{Interferometric scattering microscope}

Our microscope is configured in wide-field mode and is similar to the
setup described by Ortega-Arroyo and coworkers
\cite{ortega_arroyo_interferometric_2016}. A 450~nm, 100~mW, single-mode
diode laser (PD-01251, Lasertack) illuminates the sample. The current
driving the laser is modulated with a square wave at a frequency of
1~MHz to decrease the coherence of the laser and limit intensity
variations in the background \cite{dulin_efficient_2014}. The beam
(shown in blue in Extended Data Fig.~\ref{fig:apparatus}) is spatially
filtered by a polarization-maintaining single-mode optical fiber (fiber
1; PM-S405-XP, Thorlabs). The filtered light is collected by a lens
(lens 1; achromatic doublet, focal length = 25 mm, Thorlabs), reflected
from a polarizing beamsplitter cube (CCM1-PBS251, Thorlabs), and focused
onto the back aperture of the objective (100$\times$ oil-immersion, 1.45
NA Plan Apo $\lambda$, Nikon) to produce collimated illumination in the
imaging chamber. The light that is backscattered from the sample and
light that is reflected from the water-coverslip interface are collected
by the objective and imaged onto camera 1 (MV1-D1024E-160-CL, Photon
Focus) by the tube lens (achromatic doublet, focal length = 300 mm,
Thorlabs). We use achromatic half and quarter-wave plates (AHWP3 and
AQWP3, Bolder Vision Optik) with the polarizing beamsplitter to make an
optical isolator that minimizes the intensity lost at the beamsplitter.
The total magnification is 150$\times$, such that each pixel on the
camera views a field of 70 nm. All images are recorded with a bit depth
of 12.

The illumination intensity, set to approximately 3 kW/cm$^2$ when we
record data at 1,000 Hz and 0.3 kW/cm$^2$ at 100 Hz, is similar to that
typically used in single-molecule fluorescence
experiments\cite{moerner_methods_2003}. To minimize any possible
radiation damage, we use an exposure time that is almost equal to the
total time between frames, and we dim the imaging beam with absorptive
filters so that the camera pixels are nearly saturated. The total field
of view is 140 pixels $\times$ 140 pixels (9.8 $\upmu$m $\times$ 9.8
$\upmu$m) at 1,000 Hz and 200 pixels $\times$ 200 pixels (14 $\upmu$m
$\times$ 14 $\upmu$m) at 100 Hz.  

We use short-wavelength light ($\lambda=$~450 nm) because the intensity
of the image scales with ${\lambda}^{-2}$. While shorter wavelength
lasers are available, we find that they can damage both the sample and
optical components when used at high intensities. Control experiments at
different illumination intensities are described below and in
Supplementary Information, which also contains additional notes on the
configuration of the microscope.

The intensity of each diffraction-limited spot is
approximately linearly proportional to the number of proteins bound to
the RNA strand. The intensity of a spot is $I = I_\textnormal{r} +
I_\textnormal{s} +
2\sqrt{I_\textnormal{r}I_\textnormal{s}}\cos\phi_\textnormal{rs}$, where
$I_\textnormal{r}$ is the intensity of the reflected wave,
$I_\textnormal{s}$ the intensity of the scattered wave, and
$\phi_\textnormal{rs}$ the phase difference between the two. The term
$I_\textnormal{s}$ can be neglected since the scattered light is dim
compared to the reflected light, so the normalized intensity
$I_\textnormal{norm} = I/I_\textnormal{r}-1$ is proportional to the
total polarizability of the assembling particle
\cite{young_2018_quantitative}, which is approximately the sum of a
protein component and an RNA component. Because the RNA component is
static, it is part of the background, which is subtracted. As a result,
the normalized intensity is linearly proportional to the number of
proteins in the assembling particle.

\subsection*{Active stabilization}

The position of the coverslip relative to the objective is actively
stabilized to a few nanometers in all three dimensions. Each dimension
is controlled separately through a proportional control loop on the PC.
During each iteration of the loop, the position of the coverslip is
measured, and the voltage driving the piezoelectric actuators is
modified to keep the coverslip in its original position.

\begin{figure}[h]
	\centering
  \includegraphics{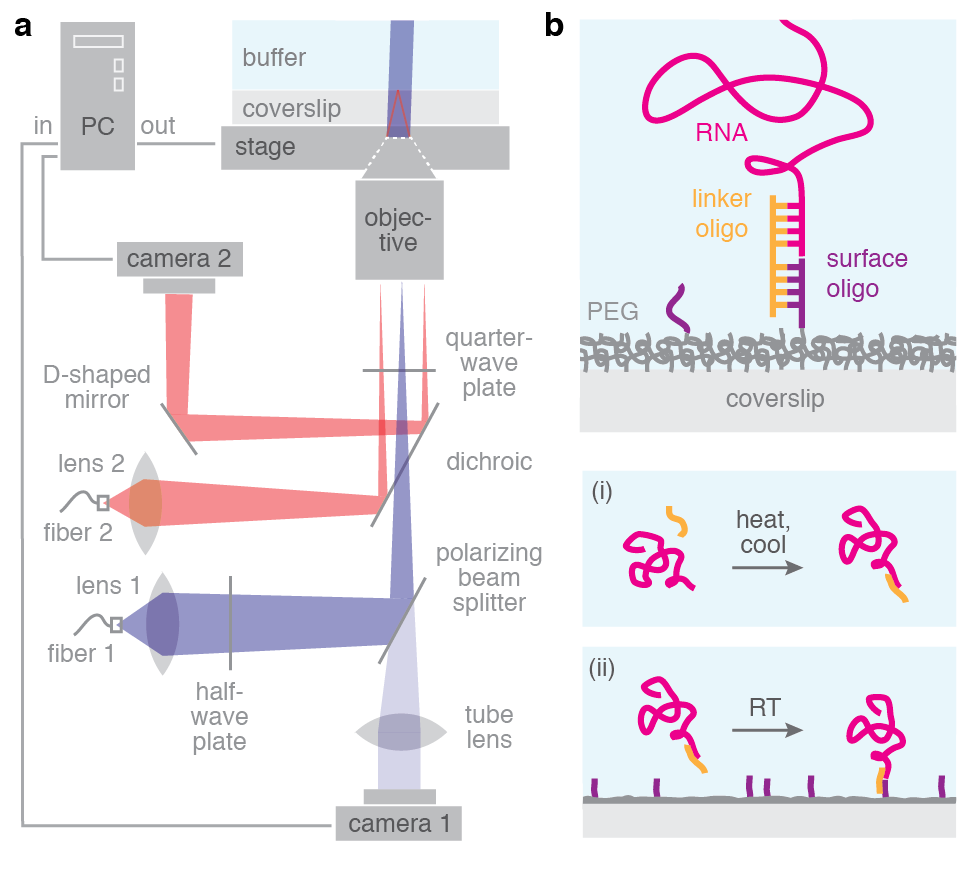}
  \captionof{figure}{\textbf{Diagram of the interferometric scattering
      microscope and DNA linkages.} (a) Our microscope is similar to the
    setup described by Ortega-Arroyo and coworkers
    \cite{ortega_arroyo_interferometric_2016}. 450-nm light (blue) is
    used for illumination. 785-nm light (red) is used for active
    stabilization in the dimension perpendicular to the coverslip
    surface. Details of the instrument are described in Methods. (b) We
    use DNA linkages~\cite{garmann_simple_2015} to bind MS2 RNA to the
    surface of a microscope coverslip. Top: diagram of the basepairing
    between the 5'-end of the RNA, a linker oligo, and a surface oligo
    that is covalently bound to the PEG-functionalized coverslip.
    Bottom: to construct the linkages we (i) bind the RNA to the linker
    oligo in solution by thermal annealing, and then (ii) add the
    RNA-DNA complexes to the functionalized coverslips at room
    temperature. Details of the process are described in Methods.}
	\label{fig:apparatus}
\end{figure}

The height of the coverslip above the objective is measured by tracking
the position of a laser (red in Extended Data Fig.~\ref{fig:apparatus})
that is totally internally reflected by the coverslip-water interface,
as described by Ortega-Arroyo and coworkers
\cite{ortega_arroyo_interferometric_2016}. We use a 785-nm, 90 mW,
single-mode diode laser (L785P090, Thorlabs) that is coupled through a
single-mode fiber (fiber 2; S630-HP, Thorlabs). The laser is driven with
a constant current (27 mA) that is well below threshold (35 mA), which
we find improves the intensity stability of the laser. After exiting the
optical fiber, the beam is collected by lens 2 (plano-convex, focal
length = 20 mm, Thorlabs), reflects from a dichroic mirror (700-nm
short-pass, Edmund Optics), and is focused onto the back aperture of the
objective. We align the beam so that after exiting the objective, it
totally internally reflects from the coverslip-water interface and
re-enters the objective. The total power incident on the coverslip is
less than 1 $\upmu$W. The return beam reflects from the coverslip and
then from a D-shaped mirror (Thorlabs) and is detected with camera 2
(DCC1545M, Thorlabs). A long-pass filter (700-nm, Thorlabs, not shown in
Extended Data Fig.~\ref{fig:apparatus}) attenuates any light from the
imaging beam that is also incident on camera 2. When the height of the
coverslip changes, the return beam is displaced laterally across camera
2, resulting in a change in the measured center-of-brightness. Under
active stabilization, any changes in the center-of-brightness are
measured and corrected every 30 ms.

The in-plane position of the coverslip is measured by tracking a 30-nm
gold particle that is adsorbed to the coverslip surface (see next
subsection for details of how we prepare the coverslips). Before each
experiment, we find one of the adsorbed gold particles by looking for
spots that have a normalized intensity of approximately 0.2. We then
move the coverslip so that the spot is near the edge of the field of
view. Using a 16 $\times$ 16-pixel region of the field of view, we
record a static background image of the coverslip with no particles
present and then move the gold particle into the center of this small
field of view. Before tracking the position of the gold particle, we
process its image in the small field of view by subtracting off the
static background, applying a bandpass filter (passing features of size
1 to 7 pixels) to smooth the image, and taking the time-median of 33
images of the particle (recorded at 33 Hz) to reduce shot noise. We then
use the program Trackpy\cite{allan_2016_60550} to locate the position of
the particle. We use this position for the active stabilization loop,
which runs once per second. The in-plane control loop frequency (1 Hz)
is lower than that of the out-of-plane control loop (33 Hz) because of
the time required to collect the median image of the particle.

The active stabilization loops are implemented in a Python script
(\url{http://github.com/manoharan-lab/camera-controller}). The same
script includes a real-time image processing routine that allows us to
see growing MS2 particles while collecting data.

\subsection*{Coverslip and gold nanoparticle functionalization}

We adapt the protocols described by Joo and Ha \cite{Joo_single_2012} to
coat glass coverslips with a layer of PEG molecules, about 1\% of which
are functionalized with short DNA oligonucleotides. We find that many
brands of \#2 coverslips are unsuitable for assembly measurements
because they have imperfections that scatter too much light. We use only
\#2 thickness, 24 mm $\times$ 60 mm rectangular glass microscope
coverslips from Globe Scientific, Inc. Details of how we functionalize
the coverslips and decorate them with 30-nm gold particles are in
Supplementary Information.

\subsection*{Flow cell design and construction}

We build chips that each contain 10 separate flow cells above a single
coverslip. Each chip consists of two sheets of cut, clear acrylic that
are sealed together and to the coverslip with melted Parafilm (Bemis).
Each flow cell has an imaging chamber that is used for the assembly
experiments, an inlet cup to hold fluid before it is introduced into the
imaging chamber of the flow cell, a short inlet chamber to connect the
inlet cup to the imaging chamber, and an outlet chamber. We use acrylic,
a hard plastic, because we find that soft materials such as
polydimethylsiloxane lead to more warping of the coverslip during
injection of the protein. A detailed description of the flow cells and
their construction is given in Supplementary Information.

\subsection*{Growth of MS2 and purification of its coat protein and RNA}

We grow wild-type MS2 by infecting liquid cultures of \textit{E. coli}
strain C3000 (a gift from Peter Stockley at the University of Leeds) and
purifying the progeny viruses following the protocols of Strauss and
Sinsheimer \cite{strauss_purification_1963}. We purify coat protein from
the virus particles following the cold acetic acid method described by
Sugiyama, Hebert, and Hartmann \cite{sugiyama_ribonucleoprotein_1967}.
We purify RNA from freshly grown MS2 virions using an RNA extraction kit
(RNeasy, Qiagen). Details about how we assess the purity of these
materials are described in Supplementary Information.

We store the purified virus particles at 4~$^{\circ}$C and discard them
after about 1 month. We store the protein at 4~$^{\circ}$C and discard
it after 1 week. We store the RNA at $-$80~$^{\circ}$C and discard it
after about 1 year.

\subsection*{Surface-immobilization of MS2 RNA by DNA linkages}
To immobilize MS2 RNA at the coverslip surface, we first hybridize the
5'-end of the RNA to a 60-base-long linker oligo (Integrated DNA
Technologies). The 40 bases at the 5'-end of the linker are
complementary to the 40 bases at the 5'-end of the RNA, and the
remaining 20 bases are complementary to the sequence of the surface
oligo (Extended Data Fig. \ref{fig:apparatus}). To anneal the linker to
the MS2 RNA, we add a 10-fold molar excess of the linker oligo to 500 nM
MS2 RNA in hybridization buffer (50 mM Tris-HCl, pH 7.0; 200 mM NaCl, 1
mM EDTA), heat the mixture to 90~$^{\circ}$C for 1~s, and then cool it
to 4~$^{\circ}$C at a rate of $-$1~$^\circ$C/s. Excess linker is removed
with a 100-kDa-MWCO centrifugal filter unit (EMD Millipore) at 14,000 g.
The 60-base-long oligonucleotides do not pass through the filter;
instead, they stick to the membrane. We confirm RNA-DNA binding by
native 1\% agarose gel electrophoresis (Supplementary Information). We
confirm that the RNA-DNA constructs specifically bind our
DNA-functionalized coverslips by interferometric scattering microscopy
(Supplementary Information). The sequence of the linker is
\texttt{5$'$-CGACAGGAAGTTGAGCAGGACCCCGAAAGGGGTCCCACCCAACCAACCAACCAACCAACC-3$'$}.

\subsection*{Calibration experiment}

\begin{figure}
\centering \includegraphics{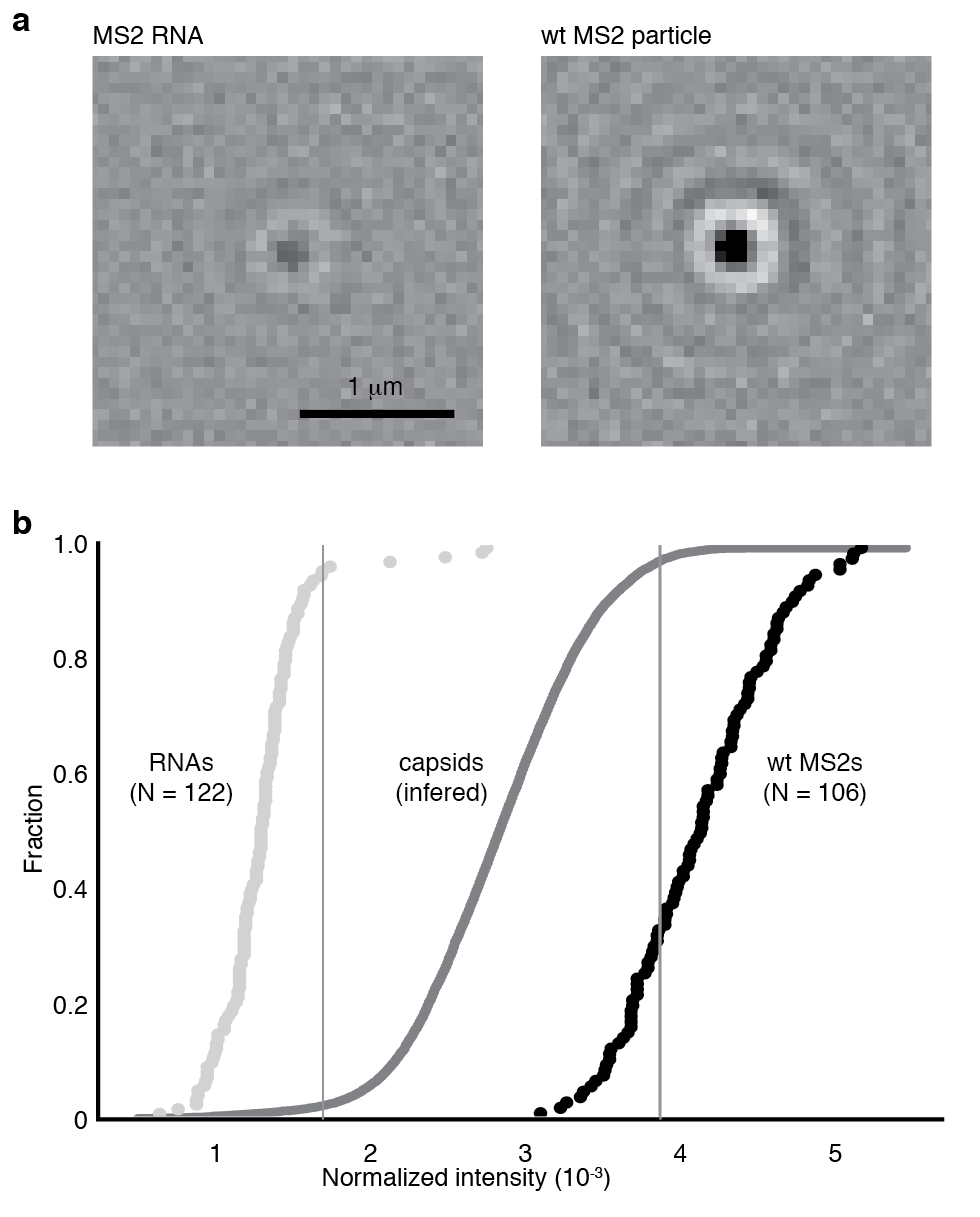}
\captionof{figure}{\textbf{Cumulative distribution of the normalized
    intensities of MS2 RNA strands and wild-type MS2 virus particles
    measured in the interferometric scattering microscope.} See Methods
  for details of the measurement. (a) Images of a single MS2 RNA strand
  (left) and a single wild-type MS2 virus particle (right). Both images
  are recorded at 100 Hz and shown with a 300-frame average. (b) We
  infer the cumulative distribution of intensities for MS2 capsids that
  fully assemble on surface-tethered RNA by convolving the intensity
  distribution of the wild-type MS2 particles with the negative of the
  intensity distribution of the MS2 RNA strands. The gray lines, which
  mark where the capsid distribution reaches 2.3\% and 97.7\%, denote
  the interval we use for identifying full capsids in the kinetic
  traces.}
	\label{fig:calibration}
\end{figure}

We measure the intensities of MS2 RNA and wild-type MS2 virus particles
(Extended Data Fig.~\ref{fig:calibration}) by imaging the particles as
they adsorb to an APTES-functionalized coverslip. For these experiments
we do not use a flow cell. Instead, we use a `lean-to' sample
chamber\cite{goldfain_dynamic_2016} made of 1-mm-thick glass slides
(Micro Slides, Corning) that are cut, cleaned by pyrolysis (PYRO-CLEAN,
Tempyrox Co.), and sealed in place with vacuum grease (High vacuum
grease, Dow Corning). To perform the calibration experiment, we first
fill the sample chamber with TNE buffer (50 mM Tris-HCl, pH 7.5; 100 mM
NaCl; 1 mM EDTA) and focus the microscope onto the coverslip. We then
exchange the buffer in the sample chamber with a solution containing
both MS2 RNA and wild-type MS2 virus particles at a concentration of 0.1
nM each in TNE buffer. We record movies (100 Hz) of these particles
nonspecifically adsorbing to the coverslip.

We see two well-separated populations in the distribution of intensities
of the particles that bind (Extended Data Fig.~\ref{fig:calibration}).
We assume that the lower-intensity population is due to the RNA strands
and the higher-intensity population is due to the MS2 viruses. To
determine the median and width of each intensity population, we separate
the two using an intensity threshold (0.003) that lies between them.
  
\subsection*{Assembly experiments}

For assembly experiments, we fill a flow cell with hybridization buffer
containing 0.2\% Tween-20 (Sigma-Aldrich) and let it sit for 10 min. We
find that this 10-min incubation with Tween-20 prevents the MS2 coat
protein from adsorbing to the coverslip through defects in the PEG
layer. Next, we flush out the Tween-20 with fresh hybridization buffer,
find the center of the imaging chamber, focus the microscope onto the
coverslip, and begin the out-of-plane active stabilization control loop.
Then we locate a 30-nm gold particle within 50 $\upmu$m of the center of
the imaging chamber and start the in-plane active stabilization control
loop. With the setup actively stabilized in all three dimensions, we
inject 1 nM RNA-DNA complexes in hybridization buffer and record a
short movie of them adsorbing to the coverslip. After 10--100 complexes
bind, we flush the imaging chamber by pumping 120~$\upmu$L of assembly
buffer through the chamber over the course of 12~min. Then we start
recording a movie and inject the coat-protein dimers in
assembly buffer. The injection starts 4~s into the movie.

\subsection*{Image processing}

We process the images to normalize them and to reduce fluctuations in
the background intensity. We apply an approach similar to the
`pseudo-flat-fielding' method described by Ortega-Arroyo and coworkers
\cite{ortega_arroyo_interferometric_2016}. The images in Fig.
\ref{fig:overview}, and Extended Data Fig. \ref{fig:calibration}, are
processed in this way, as are all the movies included in the
Supplementary Information.

Each raw image, denoted $I_\textnormal{raw}$, is processed according to
the following steps: First, a dark image, $I_\textnormal{dark}$, is
acquired by taking the time-median of many frames (200 frames for 100 Hz
data and 2,000 for 1,000 Hz data) when the illumination beam is blocked.
This image is subtracted from each raw image, yielding
$I_\textnormal{bkgd} = I_\textnormal{raw} - I_\textnormal{dark}$.
Second, features bigger than $\sigma_1 = 1.5$ pixels are removed by
subtracting a Gaussian blur, yielding $I_\textnormal{smooth} =
I_\textnormal{bkgd} - \textnormal{blur}(I_\textnormal{bkgd}, \sigma_1)$,
where $\textnormal{blur}(I, \sigma)$ is 2D Gaussian blur of the image,
$I$, using a standard deviation $\sigma$. We choose $\sigma_1 = 1.5$ to
minimize intensity changes that arise from time-varying background
fringes, even though this choice slightly decreases the normalized
intensities of the particles on the coverslip. Third, the image is
normalized to the background that has been blurred with $\sigma_2 = 20$
pixels, so that particles on the coverslip and stray fringes smaller
than $\sigma_2$ do not affect the normalization. This process yields
$I_\textnormal{norm} = (I_\textnormal{smooth}) /
\textnormal{blur}(I_\textnormal{bkgd}, \sigma_2)$. Because each image is
normalized independently of other images in the time-series,
fluctuations in the illumination intensity in time do not affect
$I_\textnormal{norm}$. Finally, all remaining static features in the
background are removed by subtracting the time-median of many frames
(300 frames for 100 Hz data and 3,000 for 1,000 Hz data) of the movie,
yielding the final processed image $I_\textnormal{final} =
I_\textnormal{norm} - I_\textnormal{norm,med}$. The noise in
$I_\textnormal{final}$ is set by shot noise for the first few seconds
after the background subtraction, but after this time, fluctuations in
the background intensity due to uncorrected mechanical drift are the
main source of measurement noise.

\subsection*{Identifying and measuring assembling particles}

To identify assembling particles, we manually locate the centers of all
dark spots that appear and are between 1 and 4 pixels across in each
processed interferometric scattering movie. We repeat this procedure
multiple times using different frames for the background subtraction to
ensure that no dark spots are missed. For each of these spots, we
measure the mean intensity in a circle of radius 1 pixel that is
centered on the particle as a function of time.  

We reject any spot that: (1) instantaneously appears in the movie,
indicating that it is from a particle that has adsorbed to the
coverslip; (2) is near the gold particle used for active stabilization
or near a defect on the coverslip that has comparable intensity (greater
than 0.1); (3) is near a particle that adsorbs to or desorbs from the
coverslip, such that its intensity is altered by the particle; (4) is so
close to another spot that the interference fringes of the two spots
overlap; (5) is near the edge of the field of view; or (6) grows at a
slow and consistent rate over the course of the measurement, consistent
with protein assembly in the absence of RNA. We describe how each of
these criteria are applied in the Supplementary Information.
    
\subsection*{Determining start and growth times}

\begin{figure}
	\centering \includegraphics{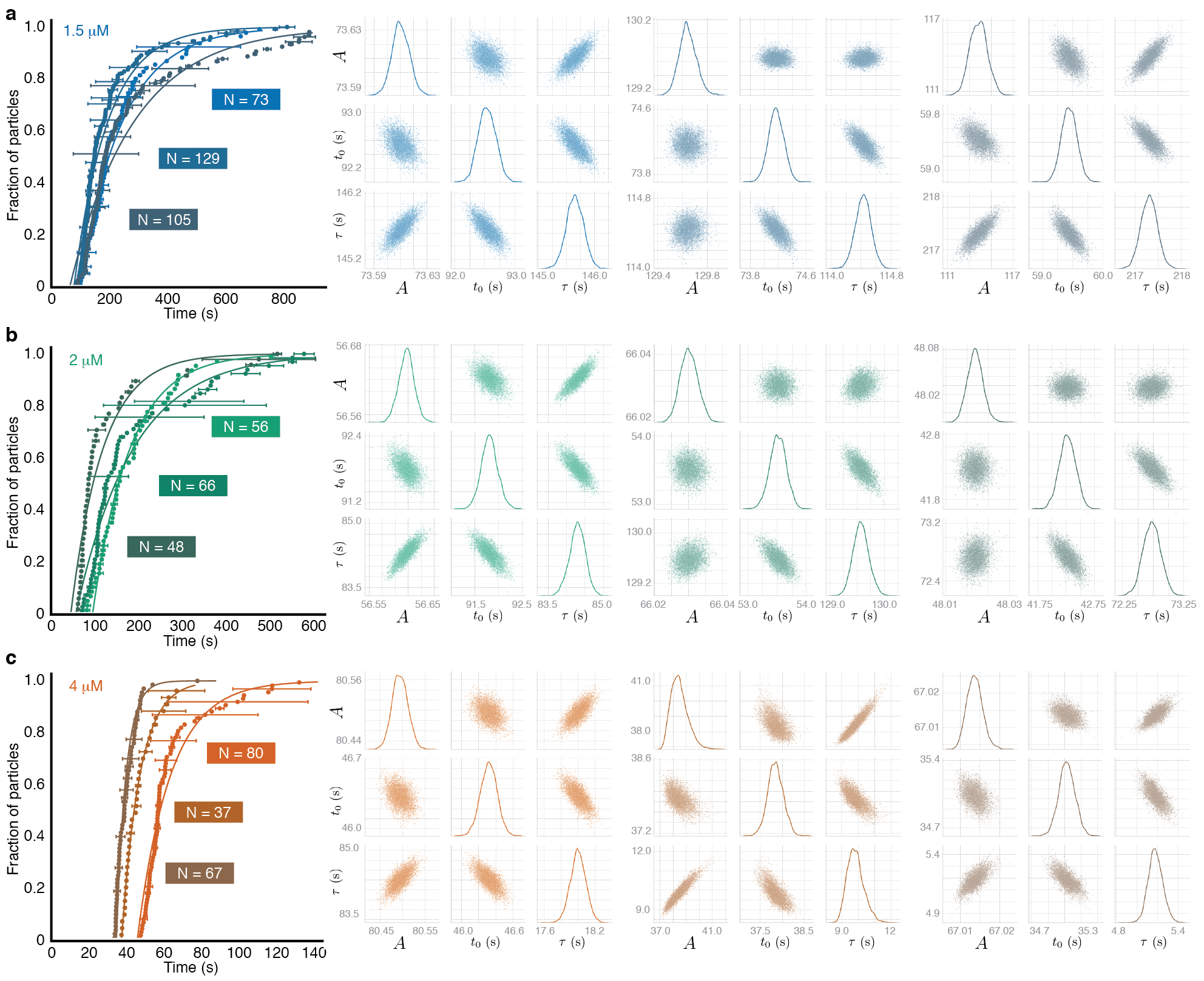}
  \captionof{figure}{\textbf{Cumulative distributions of start times and
      posterior probability distributions of parameter values obtained
      by fitting the distributions.} The results from triplicate
    assembly experiments with (a) 1.5, (b) 2, and (c) 4 $\upmu$M dimers
    are shown. Each cumulative distribution of start times (left) is
    measured from a separate assembly experiment. Uncertainties in the
    time measurements are represented by horizontal bars. Fits are shown
    as solid curves. Number of particles (N) are shown on the plot.
    Posterior probability distributions of parameter values (right) are
    sampled using a Markov-chain Monte Carlo technique. The plots along
    the diagonal show kernel density estimates of the fully marginalized
    posterior distributions of each parameter, while the off-diagonal
    plots show the joint distributions. The data and fit shown in the
    lightest color of each panel are from the experiments shown in
    Figs.~\ref{fig:MS2assembly},~\ref{fig:MS2concentration}a,
    \ref{fig:MS2concentration}c, and Extended Data
    Figs.~\ref{fig:1.5uMassembly} and \ref{fig:4uMassembly}. Data from
    all 9 of the experiments in this figure were used to obtain the
    nucleation times shown in Fig.~\ref{fig:MS2concentration}b.}
    \label{fig:nucleation_times}
\end{figure}

The cumulative distribution functions of the start times before assembly
(Figs.~\ref{fig:MS2assembly} and Extended Data
Figs.~\ref{fig:nucleation_times}) are measured as follows. Each start
time is defined as the time at which a kinetic trace reaches an
intensity of 0.001. To measure this time, we smooth each trace using a
1,000-frame moving average. The first time that the smoothed trace
reaches an intensity greater than 0.001 is called $t_1$, and the last
time that the smoothed trace has an intensity less than 0.001 is called
$t_2$ (ignoring any late detachment events or drifts in intensity). The
start time is then determined as $t_\textnormal{start} = (t_1 + t_2)/2$.
To estimate the uncertainty in each start time, we calculate the
half-width of the moving-average window and $(t_2 - t_1)/2$, and we take
the greater of the two. The cumulative distribution function of start
times is obtained by sorting the measured values of
$t_\textnormal{start}$.

We then fit the cumulative distribution to the exponential function
$N(t) = A\left(1-\exp\left[-(t-{t}_{0})/{\tau}\right]\right)$ using a
Bayesian parameter-estimation framework. A uniform, unbounded prior is
used for all parameters. The exponential function is first inverted,
yielding
\begin{equation}
  t(N) = t_0 - \tau \ln\left(1-N/A\right),
  \label{eq:model}
\end{equation}
where the fit parameters are $t_0$, $A$, and $\tau$. The posterior
probability distribution $p(t_0, A, \tau \mid D_\textnormal{CDF}, M)$,
where $D_\textnormal{CDF}$ is the observed cumulative distribution
function and $M$ is the model (Equation~(\ref{eq:model})), is then
sampled using an affine-invariant ensemble Markov-chain Monte Carlo
sampler~\cite{foreman_emcee_2013} with 50 walkers that take 500 steps
each. The walkers are initially distributed in a narrow Gaussian around
the peak of the posterior probability density function. The position of
the peak is calculated from a least-squares fit to $t(N)$. The walkers
reach an equilibrium distribution after approximately 200 steps. Pair
plots of the positions of the walkers on every step after the burn-in
are shown in Extended Data Fig.~\ref{fig:nucleation_times}, along with
the marginal distributions for each fit parameter. The best-fit
parameters reported in the text are taken as the 50\textsuperscript{th}
percentile of the marginal distributions, and the reported uncertainties
represent a credibility interval from the 16\textsuperscript{th} to the
84\textsuperscript{th} percentile.

To determine the growth time we first take the portion of each kinetic
trace that lies between the start time and the time at which the
intensity first reaches the 10\textsuperscript{th} percentile of the
capsid intensity distribution (Extended Data
Fig.~\ref{fig:calibration}), and fit this portion of the trace to a
line, using a least-squares method. We then estimate the time required
to grow a full capsid (bind 90 dimers) by approximating the growth rate
as the slope of the linear fit.

\subsection*{Control assembly experiment with lower
  illumination intensity} 

To test whether the intensity of the incident beam affects the assembly
process, we perform a set of duplicate control experiments with 2
$\upmu$M coat-protein dimers and a light intensity that is 10-fold
smaller (approximately 0.3 kW/cm$^2$). The results, shown in the
Supplementary Information, indicate that the incident light does not
qualitatively affect the assembly process.

\subsection*{TEM of assembled particles}

\begin{figure}
	\centering \includegraphics{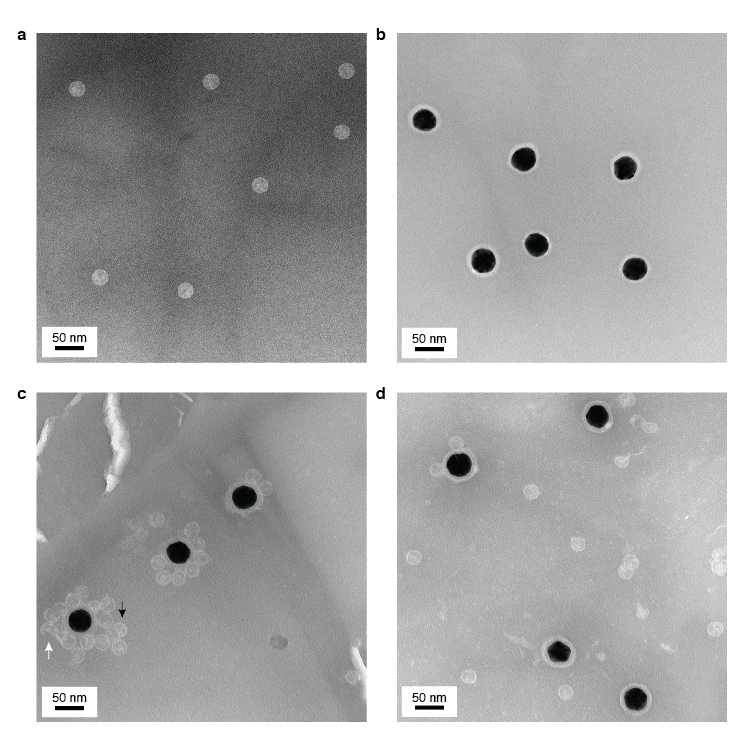}
  \captionof{figure}{\textbf{Negatively stained transmission electron
      micrographs of virus particles, functionalized gold nanoparticles,
      capsids assembled around RNA strands that are bound to the surface
      of the gold particles, and capsids assembled around untethered
      RNA.} Each sample is stained with methylamine tungstate stain
    solution (Nanoprobes) before imaging. (a) Wild-type MS2 particles.
    (b) Amine-functionalized 30-nm gold nanoparticles (Nanopartz) that
    are coated with PEG and decorated with surface oligos. The dark
    spots are the gold particles, and the surrounding lighter halos are
    the negatively stained coatings on the particle surfaces. These
    coatings consist of a proprietary polymer base layer, which is
    applied by the manufacturer to the gold nanoparticles, and the
    PEG-DNA molecules that we conjugate to the particles. (c) An
    assembly reaction in which 2 $\upmu$M coat-protein dimers in
    assembly buffer is added to RNA-DNA complexes that have been
    incubated for 1 h with the functionalized gold particles. White
    arrow points to a partial capsid. Black arrow points to a particle
    that is larger than a capsid. (d) A control reaction in which 2
    $\upmu$M coat-protein dimers in assembly buffer is added to bare RNA
    that has been incubated for 1 h with the functionalized gold
    particles. The higher number of capsids near the surface of the gold
    particles for the experiments using RNA-DNA complexes suggests that
    these capsids assembled around RNA-DNA complexes that were tethered
    to the particle surface.}
	\label{fig:TEM_tether}
\end{figure}

We use negative staining and TEM to image the protein structures that
form on MS2 RNA. First, we describe assembly experiments with MS2 RNA
that is tethered to the surface of 30-nm gold particles
(Fig.~\ref{fig:MS2assembly}e, Extended Data Fig.~\ref{fig:TEM_tether}).
The surfaces of the gold particles are functionalized in a way that is
similar to that used for the coverslips. The protocol is identical to
that used to prepare the tracer particles for active stabilization
(Supplementary Information), except that we use
NHS-PEG-$\textnormal{N}_{3}$ instead of NHS-PEG. To conjugate DNA
oligonucleotides to the PEG-coated gold particles, we add 5 $\upmu$M
DBCO-DNA to 10 nM gold particles in PBS without Ca or Mg. The mixture is
left at room temperature overnight in a tube rotator and then washed 5
times by centrifuging the mixture at 8,000 g for 5 min and resuspending
in TE buffer.

To perform the assembly reaction, we add a 100-fold molar excess of
RNA-DNA complexes (20 nM) to the gold particles (0.2 nM) and equilibrate
the mixture in TNE buffer for 1 hr on ice. We then take 6 $\upmu$L of
this mixture, add 0.42 $\upmu$L of 30 $\upmu$M coat-protein dimers
suspended in 20 mM acetic acid, and let the mixture sit for 10 min at
room temperature. The mixture is then added to a plasma-etched
carbon-coated TEM gird (Ted Pella), left to sit for 1 min, and then
removed by blotting with filter paper. Then 6 $\upmu$L of methylamine
tungstate stain solution (Nanoprobes) is added and left to sit for 1 min
before removal by blotting with filter paper. We visualize the samples
on a Tecnai F20 (FEI) transmission electron microscope operated at 120
kV. Images are captured on a 4,096 $\times$ 4,096-pixel CCD camera
(Gatan). Representative images are shown in Extended Data
Fig.~\ref{fig:TEM_tether} along with images of control reactions
involving bare RNA without the DNA linkage.

\begin{figure}
\centering
  \includegraphics{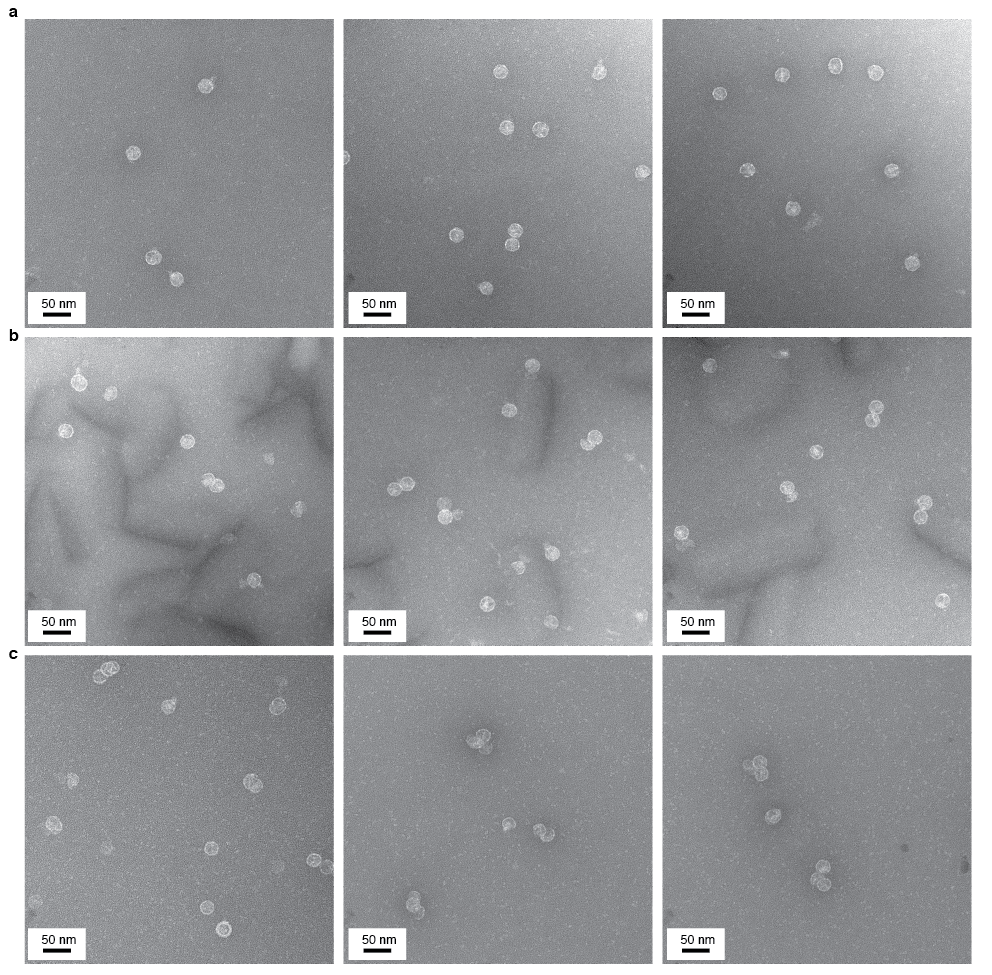}
  \captionof{figure}{\textbf{Negatively stained transmission electron
      micrographs of assembled particles from control experiments
      performed with varying concentrations of protein and untethered
      RNA.} Each sample is stained with methylamine tungstate stain
    solution (Nanoprobes) before imaging. (a) Three micrographs of
    particles taken 20 min after mixing 1.5 $\upmu$M coat-protein dimers
    and 10 nM RNA in assembly buffer. (b) Three micrographs of particles
    taken 10 min after mixing 2 $\upmu$M coat-protein dimers and 10 nM
    RNA in assembly buffer. (c) Three micrographs of particles taken 10
    min after mixing 4 $\upmu$M coat-protein dimers and 10 nM RNA in
    assembly buffer.}
	\label{fig:TEM_free}
\end{figure}

We also perform assembly reactions with RNA that is free in solution.
This is done by mixing varying concentrations of coat protein with 10~nM
of RNA in assembly buffer. After allowing the assembly reaction to
proceed for a fixed amount of time, the mixture is imaged by TEM, as
described above. Representative electron micrographs of particles
assembled with 1.5, 2, and 4~$\upmu$M coat-protein dimers are shown
in Extended Data Fig.~\ref{fig:TEM_free}.

\subsection*{Buffer recipes}

\begin{description}
\item[Assembly buffer] 42 mM Tris-HCl, pH 7.5; 84 mM NaCl; 3 mM acetic
  acid; 1 mM EDTA
\item[Hybridization buffer] 50 mM Tris-HCl, pH 7.0; 200 mM NaCl; 1 mM EDTA 
\item[TAE buffer] 40 mM Tris-acetic acid, pH 8.3; 1 mM EDTA
\item[TNE buffer] 50 mM Tris-HCl, pH 7.5; 100 mM NaCl; 1 mM EDTA 
\item[TE buffer] 10 mM Tris-HCl, pH 7.5; 1 mM EDTA
\end{description}

\subsection*{Code availability}

The code used to analyze the data is available from the corresponding
author on reasonable request.

\subsection*{Data availability}

The datasets generated and analyzed during the current study are
available from the corresponding author on reasonable request.
 
\section*{Acknowledgements}

  We thank Peter Stockley and Amy Barker at the University of Leeds for
  sending us initial stocks of MS2 virus and C3000 cells and their
  growth protocols. We thank Philip Kukura, Marek Piliarik, Vahid
  Sandoghdar, and Michael Brenner for helpful
  discussions. This work is supported by the Harvard MRSEC under
  National Science Foundation grant no.\ DMR-1420570. Additionally, this
  material is based upon work supported in part by the National Science
  Foundation Graduate Research Fellowship under grant no.\ DGE-1144152.
  It was performed in part at the Center for Nanoscale Systems (CNS), a
  member of the National Nanotechnology Coordinated Infrastructure
  Network (NNCI), which is supported by the National Science Foundation
  under NSF award no.\ 1541959. CNS is part of Harvard University.
 
\section*{Author contributions}
VNM came up with the idea of studying virus self-assembly using
interferometric scattering microscopy and developed this idea with RFG
and AMG. AMG and RFG designed the experimental setup, performed the
experiments, and analyzed the data. VNM supervised the project. AMG,
RFG, and VNM wrote the manuscript.

\bibliographystyle{achemso}
\bibliography{virus-assembly}

\clearpage

\beginsupplement

\section*{Extended Data}

\begin{figure}
	\centering
  \includegraphics{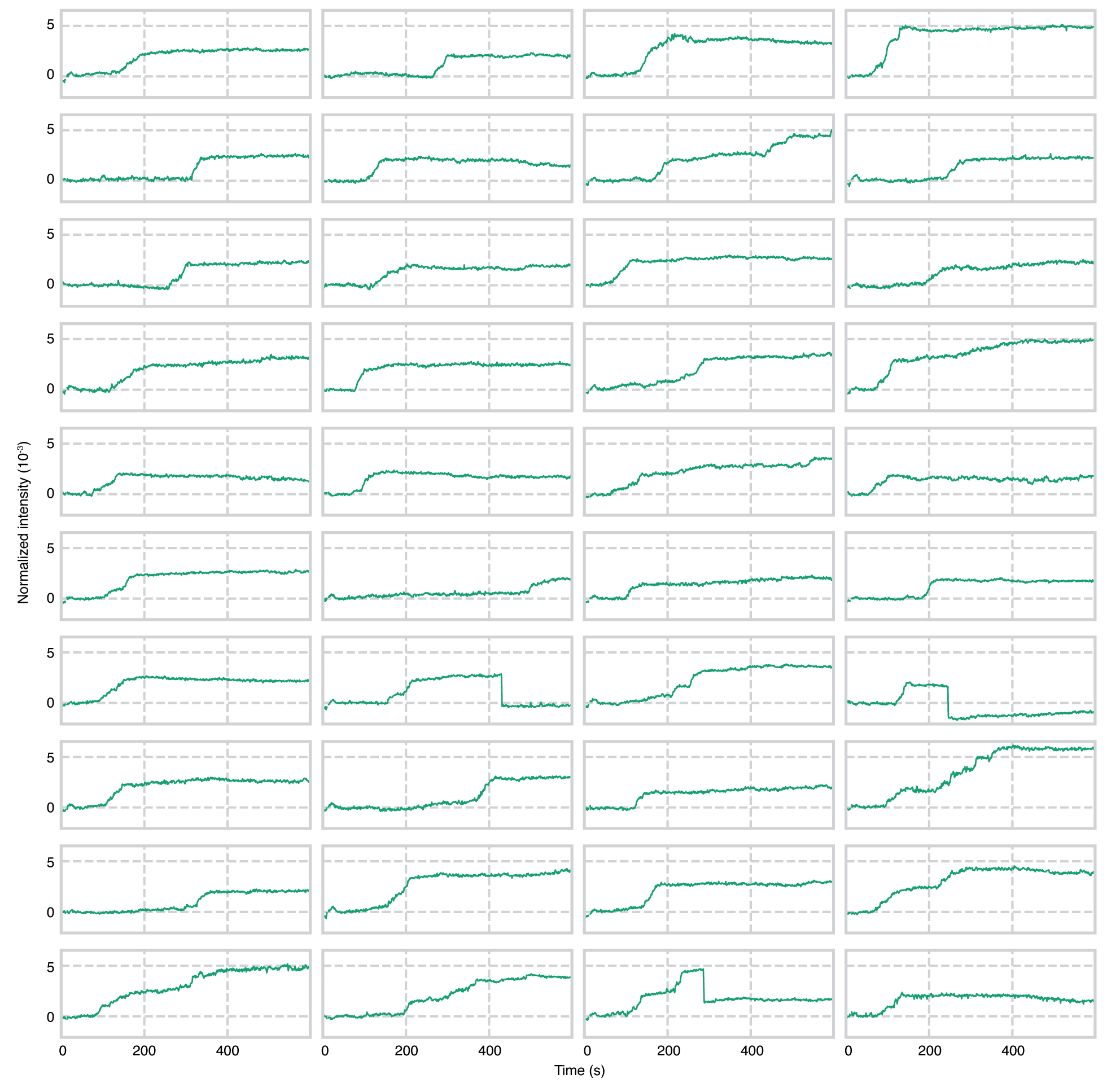}
  \captionof{figure}{\textbf{Assembly of 2 $\upmu$M coat-protein
      dimers.} Kinetic traces for 40 of the 56 observed assembling
    particles in one experiment are shown. Some traces show abrupt drops
    in intensity, which we interpret as detachment events. One of the
    above traces drops to an intensity of between -0.001 and -0.002,
    which is approximately the negative intensity of the RNA in the
    background image. We therefore interpret this event as the
    detachment of the RNA and assembled proteins from the surface. One
    trace drops to an intensity near 0, suggesting that the assembled
    protein has detached from the RNA, while the RNA remains on the
    surface. One of the traces drops from an intensity near 0.005 by an
    amount (0.0032) that corresponds to a full capsid, suggesting that
    overgrown particles can contain capsids. A portion of the traces
    from the same experiment appear in Fig.~\ref{fig:MS2assembly}, and
    one trace from the experiment appears in Fig.~\ref{fig:overview}d.
    The final intensities of the particles in this experiment are used
    for Fig.~\ref{fig:MS2concentration}c. The traces are measured from
    the data shown in Supplementary Movie~1. The data are recorded at
    1,000 Hz and are plotted with a 1,000-frame average.}
	\label{fig:2uMassembly}
\end{figure}

\begin{figure}
	\centering
  \includegraphics{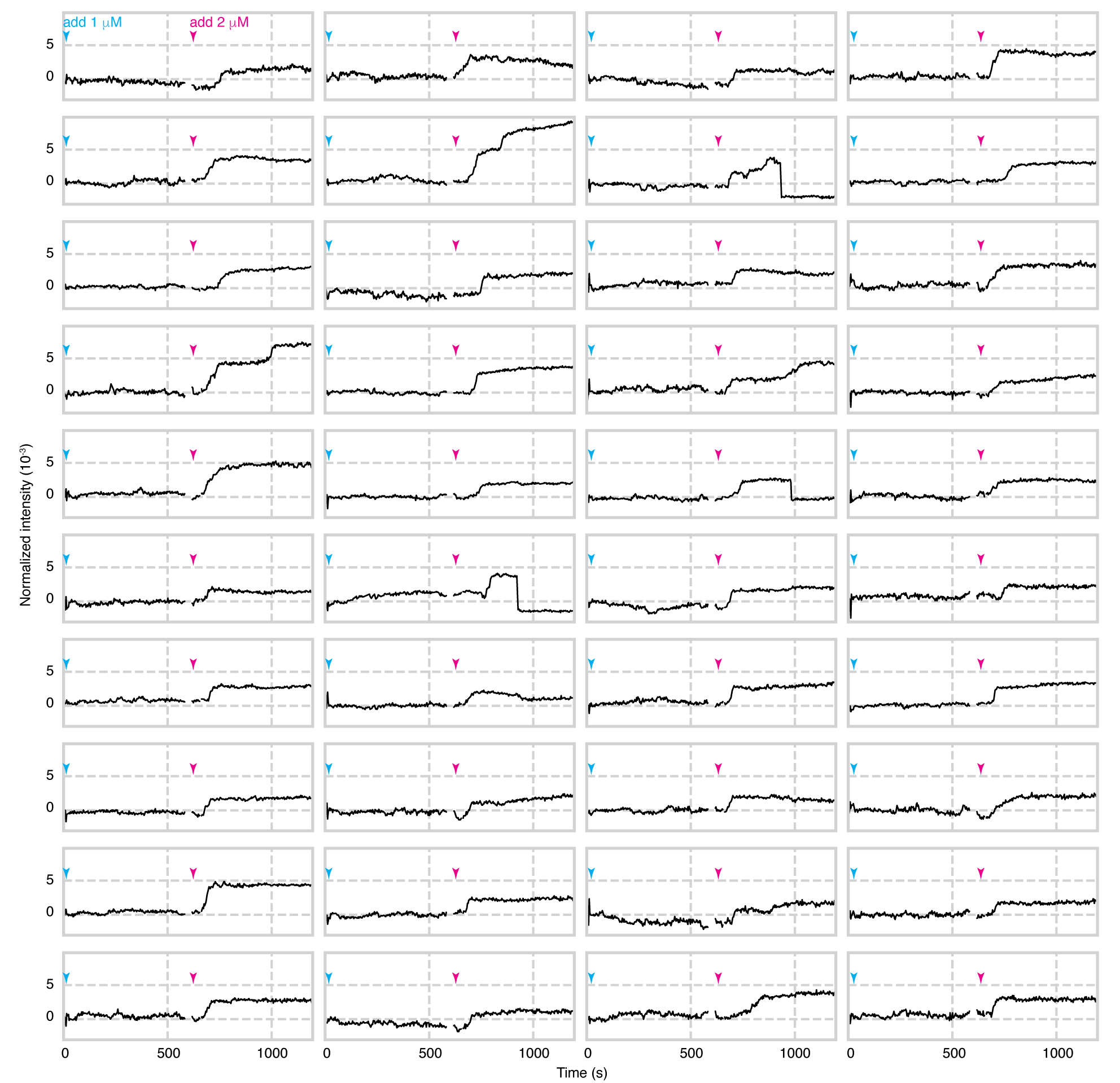}
  \captionof{figure}{\textbf{Assembly of 1 $\upmu$M coat-protein
      dimers.} When 1 $\upmu$M coat-protein dimers is added (cyan
    arrowheads) to the surface-bound RNA, no assembling particles appear
    over the course of 600~s. At this point, 2 $\upmu$M coat-protein
    dimers is added (pink arrowheads), after which we observe particles
    assembling at 75 locations within the field of view. Intensity
    traces for 40 of these particles are shown above. We also show
    traces for the first 600~s at the same locations. There is no data
    between 586 and 615 s, during which time we block the illumination
    beam and inject the 2 $\upmu$M protein. As in Extended Data
    Fig.~\ref{fig:2uMassembly}, we interpret abrupt drops in intensity
    after assembly as detachment events. The traces are measured from
    the data shown in Supplementary Movie~2. The data are recorded at 100
    Hz and are plotted with a 300-frame average.}
	\label{fig:1uMassembly}
\end{figure}

\begin{figure}
\centering
  \includegraphics{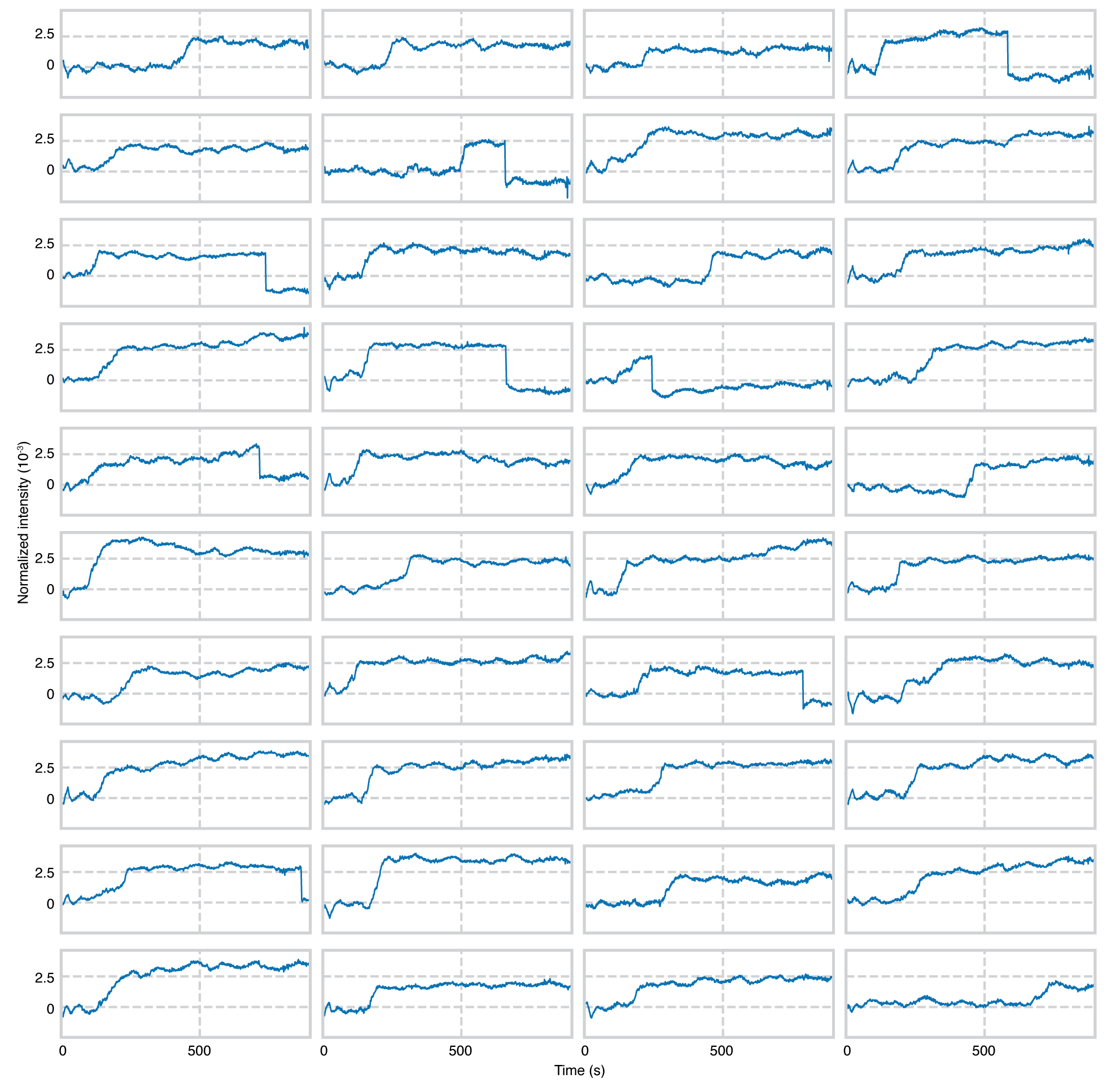}
  \captionof{figure}{\textbf{Assembly of 1.5 $\upmu$M coat-protein
      dimers.} Kinetic traces for 40 of the 73 observed assembling
    particles in one experiment are shown. As in Extended Data
    Fig.~\ref{fig:2uMassembly}, we interpret abrupt drops in intensity
    after assembly as detachment events. A portion of the traces from
    the same experiment appear in Fig.~\ref{fig:MS2concentration}. The
    final intensities of the particles in this experiment are used for
    Fig.~\ref{fig:MS2concentration}c. The traces are measured from the
    data shown in Supplementary Movie~3. The data are recorded at 1,000
    Hz and are plotted with a 1,000-frame average.}
	\label{fig:1.5uMassembly}
\end{figure} 

\begin{figure}
\centering
  \includegraphics{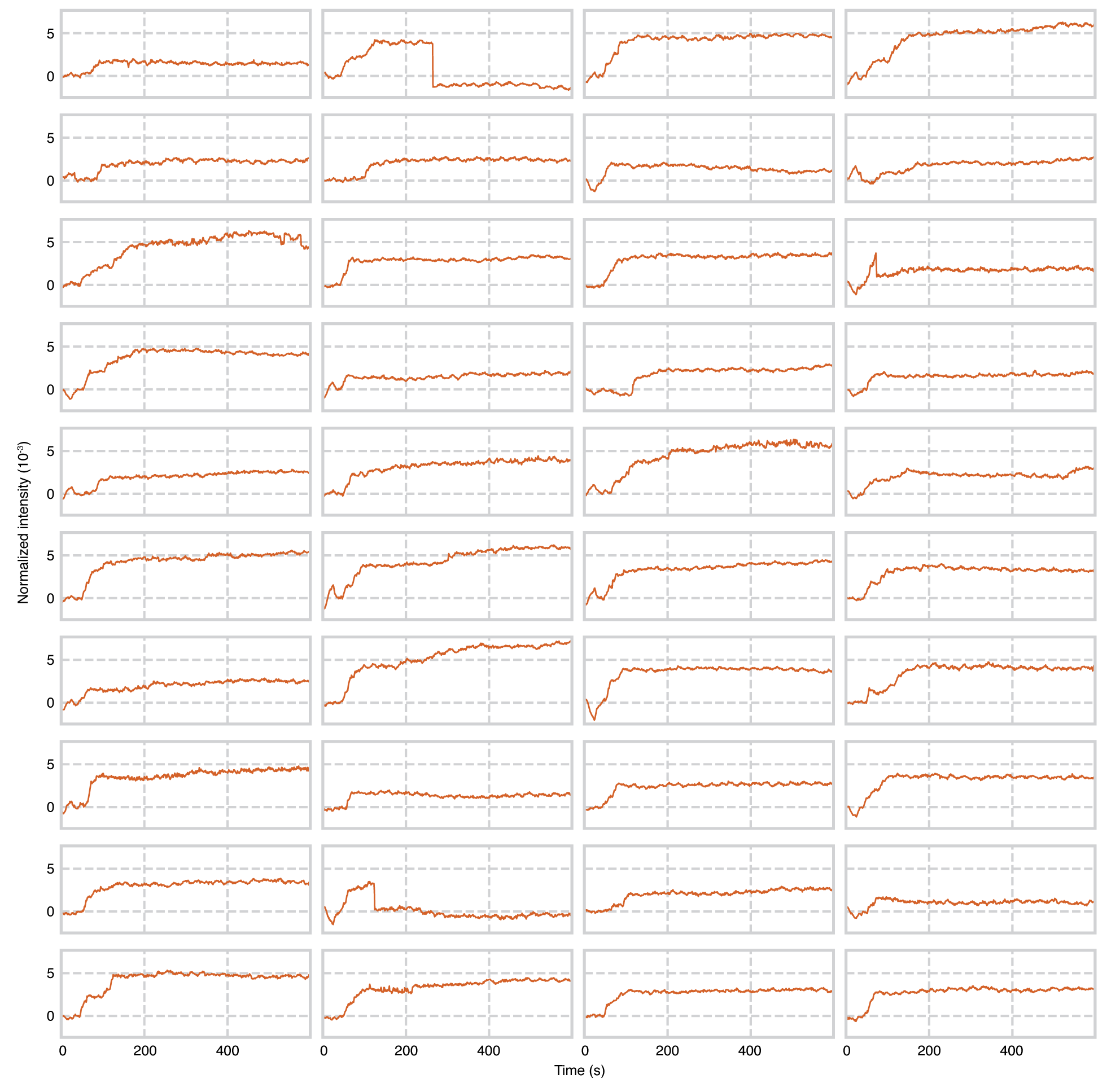}
  \captionof{figure}{\textbf{Assembly with 4~$\upmu$M coat-protein
      dimers.} Kinetic traces for 40 of the 80 observed assembling
    particles in one experiment are shown. As in Extended Data
    Fig.~\ref{fig:2uMassembly}, we interpret abrupt drops in intensity
    after assembly as detachment events. A portion of the traces from
    the same experiment appear in Fig.~\ref{fig:MS2concentration}. The
    final intensities of the particles in this experiment are used for
    Fig.~\ref{fig:MS2concentration}c. The traces are measured from the
    data shown in Supplementary Movie~4. The data are recorded at 1,000
    Hz and are plotted with a 1,000-frame average.}
	\label{fig:4uMassembly}
\end{figure}

\end{document}


\maketitle

\renewcommand{\figurename}{\textbf{Supplementary Figure}}

\section*{Time for protein to reach the surface-bound
  RNA}
Here we estimate how long it takes MS2 coat-protein dimers to reach the
surface-bound RNA molecules after the protein is pumped into the imaging
chamber. This time scale is set by the rate of diffusion and the
distance between the protein and coverglass when it is first introduced.

We first model how fluid is introduced into the imaging chamber.
Downstream of the inlet cup, the flow cell contains a cylindrical inlet
chamber (1~mm diameter, 3~mm long), which is followed by the imaging
chamber (0.75~mm tall, 1.0~mm wide, and 4.6~mm long) that contains our
field of view. The field of view is in the center of the bottom surface
of the imaging chamber. To simplify our calculations, we assume that the
flow cell consists of a single cylindrical chamber with a radius $R =
0.375$~mm and that our field of view is $L=9.3$~mm from the entrance to
the cylinder. The diameter of the cylinder is chosen to match the height
of the imaging chamber, and the length $L$ is chosen so that the volume
$\pi LR^2$ is the same as the total volume in the actual inlet and
imaging chambers upstream of the field of view.

\begin{figure}[h]
	\centering
  \includegraphics{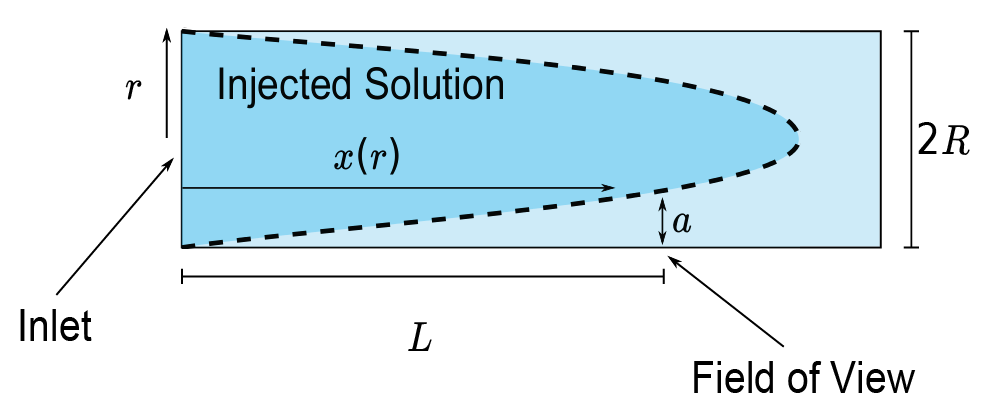}
  \caption{\textbf{Model of the flow profile for the injected protein.}
    We model the flow chamber as a cylinder, as discussed in Methods.
    The dashed line represents the parabolic boundary between the
    injected protein solution and the solution that is already in the
    chamber.}
	\label{fig:flowmodel}
\end{figure}

We assume a no-slip boundary condition, such that the flow profile in
the model cylindrical chamber is laminar and parabolic
\cite{brenner_hydrodynamics_1983}. In our experiments, we inject $V=$~10
$\upmu$L of fluid over 20 s, so that the average flow velocity is
approximately 0.5 mm/s, yielding a Reynolds number of 0.5, which
justifies the laminar assumption. We further assume that the diffusion
of protein across the chamber is negligible over the duration of the
pumping, so that the parabolic front that separates the new protein
solution from the old buffer solution is sharply defined. Indeed, the
time for an MS2 coat-protein dimer (hydrodynamic radius 2.5 nm
(Ref.~\citenum{borodavka_evidence_2012}) with diffusion coefficient,
$D=$~90 $\upmu$m$^2$/s) to diffuse across the cylinder radius is
approximately 1,600~s, much longer than the pumping duration. The shape
of the parabolic boundary is described by $x(r) = \left(2V/\pi
  R^2\right)(1-(r/R)^2)$, where $r$ is the radial coordinate of the
cylinder, and $x(r)$ is the distance down the cylinder from the end
where the protein is injected (Supplementary
Figure~\ref{fig:flowmodel}). Note that in the center of the cylinder,
$x(r = 0) \approx$~45 mm. Thus, the tip of the parabola following a pump
of V = 10~$\upmu$L extends well beyond the field of view. Above the
field of view, the distance from the parabolic boundary to the surface
is $a = R(1 - \sqrt{1-\pi R^2L/2V} ) \approx$~40 $\upmu$m. This is the
distance that the protein must diffuse to reach the surface-bound RNA.

To experimentally determine the distance from the parabolic boundary to
the surface just after the pump, we use a bright-field microscope
(Eclipse Ti, Nikon) and tracer particles (1 $\upmu$m sulfate-latex,
Invitrogen). For this experiment we fill the flow cell with water,
position our field of view in the center of the imaging chamber, inject
a solution of tracer particles (0.08\% w/v in water), and measure the
distance of the tracer particles from the coverslip immediately after
the injection. We find that there is a well-defined boundary between the
solutions with and without particles, and that this boundary is $a =$
20--50~$\upmu$m above the coverslip, depending on the pump and flow cell
used. This distance agrees well with the distance calculated above
($40~\upmu$m).

With this length scale and the diffusion coefficient, we can calculate
the time it takes proteins to diffuse to the surface, $t_\textnormal{D}
= a^2/D$. More specifically, at a distance $a$ from the protein
solution, $t_\textnormal{D}$ is the time it takes for the concentration
of proteins to reach approximately half the injected concentration. We
find that, for a 20--50 $\upmu$m distance, $t_\textnormal{D}=$~5--30 s.

This timescale agrees with the measured delay time that precedes
assembly in our experiments. For our assembly experiments, we stop
pumping 24~s after the time-series begins, so the concentration of
protein at the surface should reach half of the injected concentration
about 30--55~s after the beginning of the time-series. Since we do not
observe assembly on our experimental time scales when introducing
1~$\upmu$M protein dimers (Extended Data Figure~6), we do not
expect to detect assembling particles until after the concentration of
dimers at the surface exceeds 1~$\upmu$M. Thus, when introducing
1.5~$\upmu$M dimers, we do not expect to detect assembling particles
until after this 30--55~s delay, which is indeed the case (Extended Data
Figure~5). Also, when introducing 2~$\upmu$M and 4~$\upmu$M protein
dimers, we expect to detect assembling particles a bit sooner but not
before this 30--55~s delay, which is again what we observe (Extended
Data Figure~5).

\section*{Configuration of the interferometric microscope}

In our microscope, the imaging beam is slightly misaligned to reduce
back-reflections from the objective and the roof of the imaging chamber.
To keep the point-spread function of the microscope symmetric, we set
the misalignment as small as possible such that back-reflections from
the objective do not overlap with the reference beam on the camera. To
accomplish this, we first align the imaging beam with the microscope
axis, and then we offset fiber 1 laterally using a two-axis linear
translation stage (Thorlabs) and tilt the imaging beam using a mirror
mounted in a two-axis kinematic mount (Thorlabs) located between lens 1
and the half-wave plate.

To minimize vibrations and long-term mechanical drift, we make the
imaging beam path as short as possible, we mount the apparatus on an
isolated optical table (RS4000, Newport), and we secure all cables going
to non-isolated equipment using clamps that we line with semi-rigid foam
(0.75-in-thick polyethylene; 8865K522, McMaster-Carr). To minimize
thermal drift and the effects of air currents, we cover the entire
apparatus in a foam-core box. We also allow all electronics associated
with the microscope to warm up for a few hours before starting an
experiment, so that any thermal gradients can equilibrate.

The coverslip and flow cells are mounted on a motorized three-axis stage
(MAX343, Thorlabs) that has stepper motors for coarse adjustments and
piezoelectric actuators for fine adjustments. The fine adjustments are
used for active stabilization. 

\section*{Coverslip functionalization}

We treat the coverslips with (3-amino\-pro\-pyl)\-tri\-ethoxy\-silane
(APTES, 99\%, Sigma-Aldrich) to impart a positive surface charge when
the coverslips are submerged in neutral-pH buffer. The coverslips can
then nonspecifically bind oppositely charged macroions such as nucleic
acids and MS2 capsids, as shown in Extended Data Figure~3. Furthermore,
the layer of amino groups can form covalent linkages through
\textit{N}-hydroxysuccinimide (NHS) chemistry. We form the PEG layer by
adding 90-$\upmu$L of 100 mM sodium bicarbonate buffer containing 9 mg
of a 100:1 mixture of 5,000-Da NHS-PEG (> 95\%, Nanocs) and 5,000-Da
NHS-PEG-N$_{3}$ (purity unreported, Nanocs) between two
APTES-functionalized coverslips and then letting the `sandwich' sit
overnight at room temperature in a humid box before washing the slips
with deionized water (obtained from a Millipore RNase-free system;
Synthesis, Milli-Q). We attach DNA oligonucleotides to the surface-bound
NHS-PEG-N$_{3}$ molecules by copper-free click chemistry. The
20-base-long oligonucleotides are synthesized with a dibenzocyclooctyne
(DBCO) group on the 5'-end (RNase-free HPLC purified, Integrated DNA
Technologies). We place 90 $\upmu$L of 10 $\upmu$M DBCO-DNA in
phosphate-buffered saline (PBS without Ca or Mg, Lonza) between two
coverslips and let the sandwich sit overnight at room temperature in a
humid box. The sequence of the surface oligo is
\texttt{5'-(DBCO)-GGTTGGTTGGTTGGTTGGTT-3'}.

\begin{figure}
	\centering
  \includegraphics{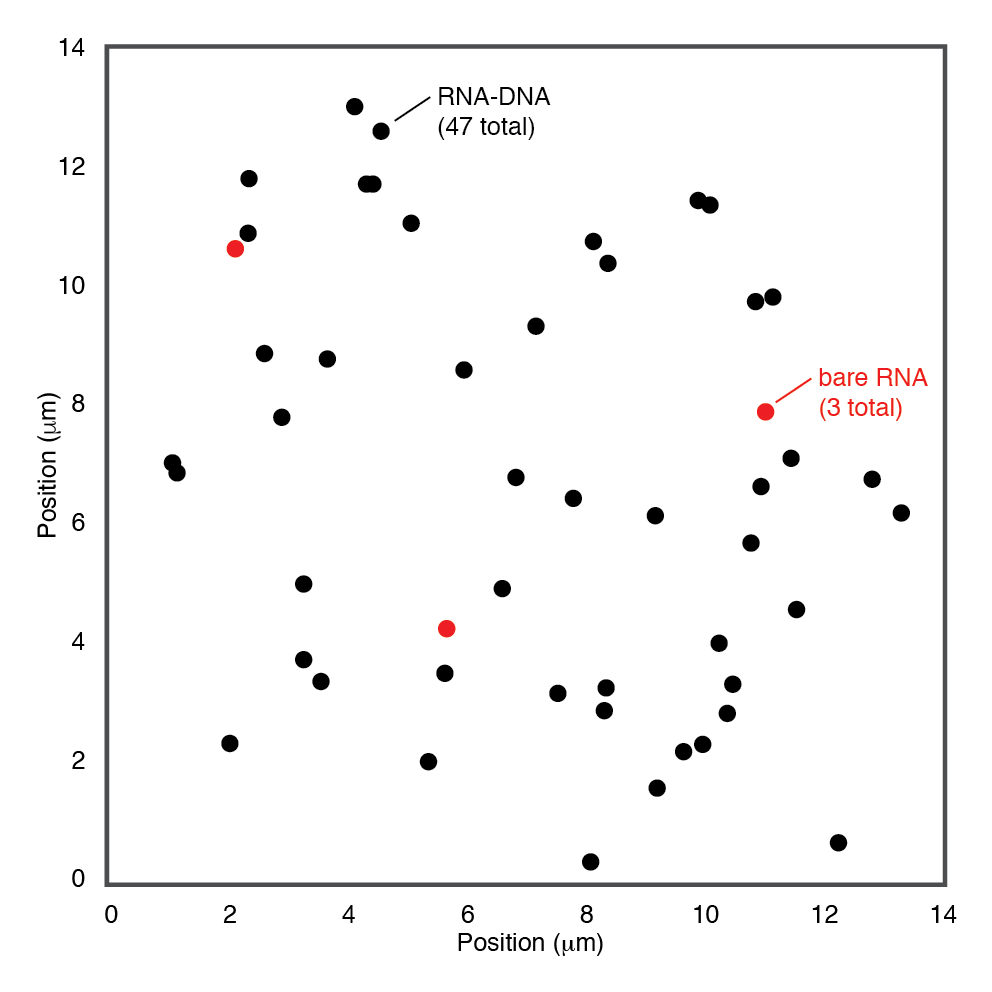}
  \caption{\textbf{Specific binding of RNA to the coverslip
      via DNA linkages.}
    To test whether the DNA-linkage enables specific binding of the RNA
    to the coverslip, solutions containing 1 nM of either bare RNA or
    RNA-DNA complexes in hybridization buffer are injected into the
    imaging chamber of the interferometric scattering microscope. If the
    binding is specific, we expect only the RNA-DNA complexes to stick
    to the coverslip surface. The bare RNA is injected first, and we
    image the system for 60~s to detect each molecule that binds. We
    then inject the RNA-DNA complexes, and we repeat the measurement.
    The location of each detected binding event is shown: we observe a
    total of 3 bare RNA molecules (red circles) and 47 RNA-DNA complexes
    (black circles). We conclude that the binding between the RNA-DNA
    complexes and the coverslip is highly specific, and that most of the
    RNA-DNA complexes that are bound to the coverslip are tethered by a
    DNA linkage.}
	\label{fig:RNAbinding}
\end{figure}

We test that the coverslips are functionalized with the surface oligo by
measuring the specific binding of MS2 RNA that is hybridized to a
complementary linker oligo using interferometric scattering microscopy
(Supplementary Figure~\ref{fig:RNAbinding}).

Following functionalization, we decorate the coverslips with 30-nm gold
particles that serve as tracer particles for active stabilization. We
purchase 30-nm amine-functionalized particles (Na\-no\-partz) and
conjugate them to NHS-PEG to prevent adsorption of coat proteins. The
conjugation is done by adding 20 mg of NHS-PEG to 200 $\upmu$L of 10 nM
of gold particles in 100 mM sodium bicarbonate buffer. The mixture is
left overnight in a tube rotator. The particles are then washed five
times by centrifuging the mixture at 8,000~g for 5 min and then
resuspending in TE buffer (10 mM tris-HCl, pH 7.5; 1 mM EDTA). To allow
the the PEG-passivated gold particles to bind non-specifically to the
coverslip, we sandwich 70 $\upmu$L of 0.1~nM suspension of the particles
between two coverslips and let them sit for 10 min at room temperature
before washing the slips with deionized water. The method produces an
average surface density of about 1 particle per 100 $\upmu$m$^{2}$, as
measured in the interferometric scattering microscope. Functionalized
coverslips are stored under nitrogen gas at $-$20~$^{\circ}$C and
discarded after 2 months.

\section*{Flow cell design and construction}

A schematic of each of the layers of the chip, a cross sectional view of
a single flow cell, and a photograph of the final assembled state are
shown in Supplementary Figure~\ref{fig:flowcell}. The bottom acrylic
sheet (Optix Acrylic, ePlastics) is 0.75 mm thick and contains 10
rectangular through-holes (1~mm~$\times$ 4.6~mm) that are cut with a
laser cutter (HSE 150W, KERN). These rectangular holes form the imaging
chamber of each flow cell. The top acrylic sheet (6.35~mm thick cast
acrylic, McMaster-Carr) serves as the roof of the imaging chambers and
contains the inlet cups, the inlet chambers, and the outlet chambers.
Each inlet cup is 3.35~mm deep and 4 mm in diameter. Each inlet chamber
is a 1-mm-diameter through-hole that begins at the base of an inlet cup
and connects to an imaging chamber in the bottom acrylic sheet. The
outlet chambers are 1.6-mm-diameter through-holes. We epoxy (5 minute
epoxy, Devcon) a 10-mm-long aluminum tube (inner diameter 0.9~mm, outer
diameter 1.6~mm, McMaster-Carr) into each outlet chamber. All holes in
the top acrylic piece are machined with a mill. The Parafilm sheets used
to seal together the layers of the flow cell contain rectangular gaps
that are the same size as the imaging chambers. The gaps are cut with a
computer-controlled vinyl cutter (CAMM-1 Servo, Roland).

\begin{figure}
  \begin{adjustwidth}{-1in}{-1in}
  \centering
  \includegraphics{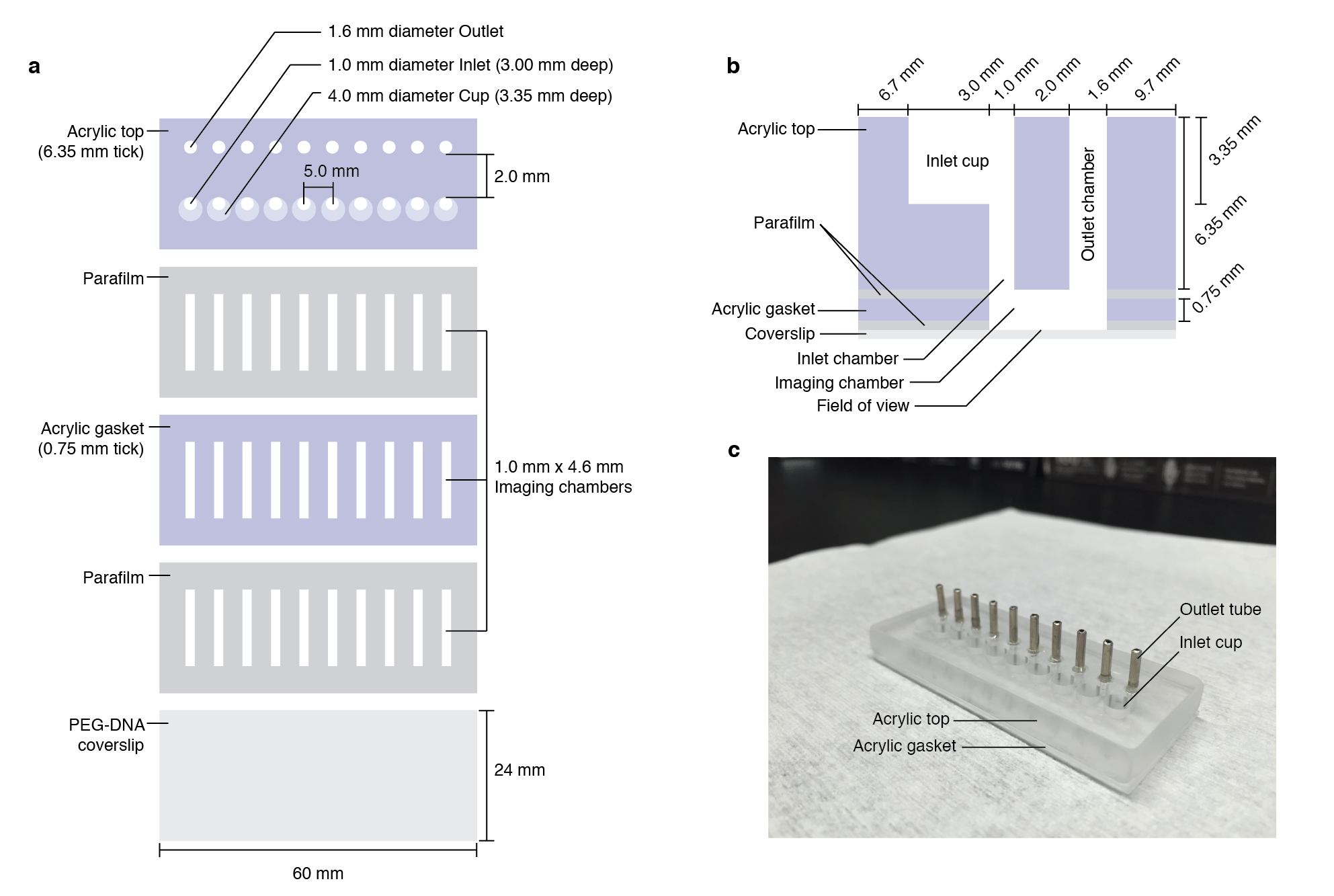}
  \caption{\textbf{A schematic of the rigid flow cell.} (a)
    Each of the layers used to build the flow cell are stacked on top of
    one another. When heated, the Parafilm seals the layers together.
    (b) A cross-section of a flow cell. (c) A photo of an assembled flow
    cell. Aluminum tubes are epoxied into the outlet chambers to connect
    to the Tygon tubing.}
	\label{fig:flowcell}
  \end{adjustwidth}
\end{figure}

To assemble each chip, we first clean the acrylic sheets and Parafilm by
sonicating in a 2\% w/v aqueous solution of sodium dodecyl sulfate
(>99\%, Sigma-Aldrich) for 30 min. After sonicating, we rinse the
acrylic and Parafilm with deionized water and then dry them under a
stream of nitrogen gas. Next, we press one sheet of Parafilm onto the
bottom acrylic sheet so that the Parafilm and acrylic stick together,
and we place this assembly in a 65~$^{\circ}$C oven for 5 min. The top
acrylic sheet is also placed in the oven for 5 min. When we remove the
acrylic sheets from the oven and press them firmly together, the melted
Parafilm seals the two sheets of acrylic together to form the chip. We
then press the other sheet of Parafilm onto the bottom of the chip so
that it sticks, and we place the chip in a 65~$^{\circ}$C oven for 5
min. We remove the chip from the oven and press it firmly onto the
functionalized coverslip (which is not heated) to seal the chip to the
coverslip. We use all of the flow cells on a coverslip within one or two
days.

We inject buffer solution into each imaging chamber using a plastic
syringe (3 mL BD, VWR) that is connected to the aluminum outlet tube by
a short (approximately 4 cm) length of tubing (Tygon PVC,
McMaster-Carr). We fill the inlet cup with solution and then pull it
through the imaging chamber by actuating the syringe with a motorized
linear translation stage (PT1-Z8, Thorlabs). Each time we inject a
solution into the imaging chamber, we use the motorized stage to inject
10 $\upmu$L of solution at a constant rate over 20 s. Before further
injections we use a Kimwipe (Kimberly-Clark Professional) to wick any
remaining solution from the inlet cup. To ensure that the fluid
injection is reproducible, we prevent any air bubbles from entering the
flow cell, tubing, or syringe. We mount the syringe vertically to
prevent air bubbles from being trapped inside it.

\section*{Assessing the concentration and purity of MS2 virus particles, and of isolated coat protein and RNA} 

We determine the concentration of MS2 by UV-spec\-tro\-photo\-metry
(Nano\-Drop-1000, Thermo Scientific), assuming an extinction coefficient
of 8.03 mL~mg$^{-1}$~cm$^{-1}$ at 260 nm
(Ref.~\citenum{strauss_purification_1963}).

\begin{figure}
	\includegraphics{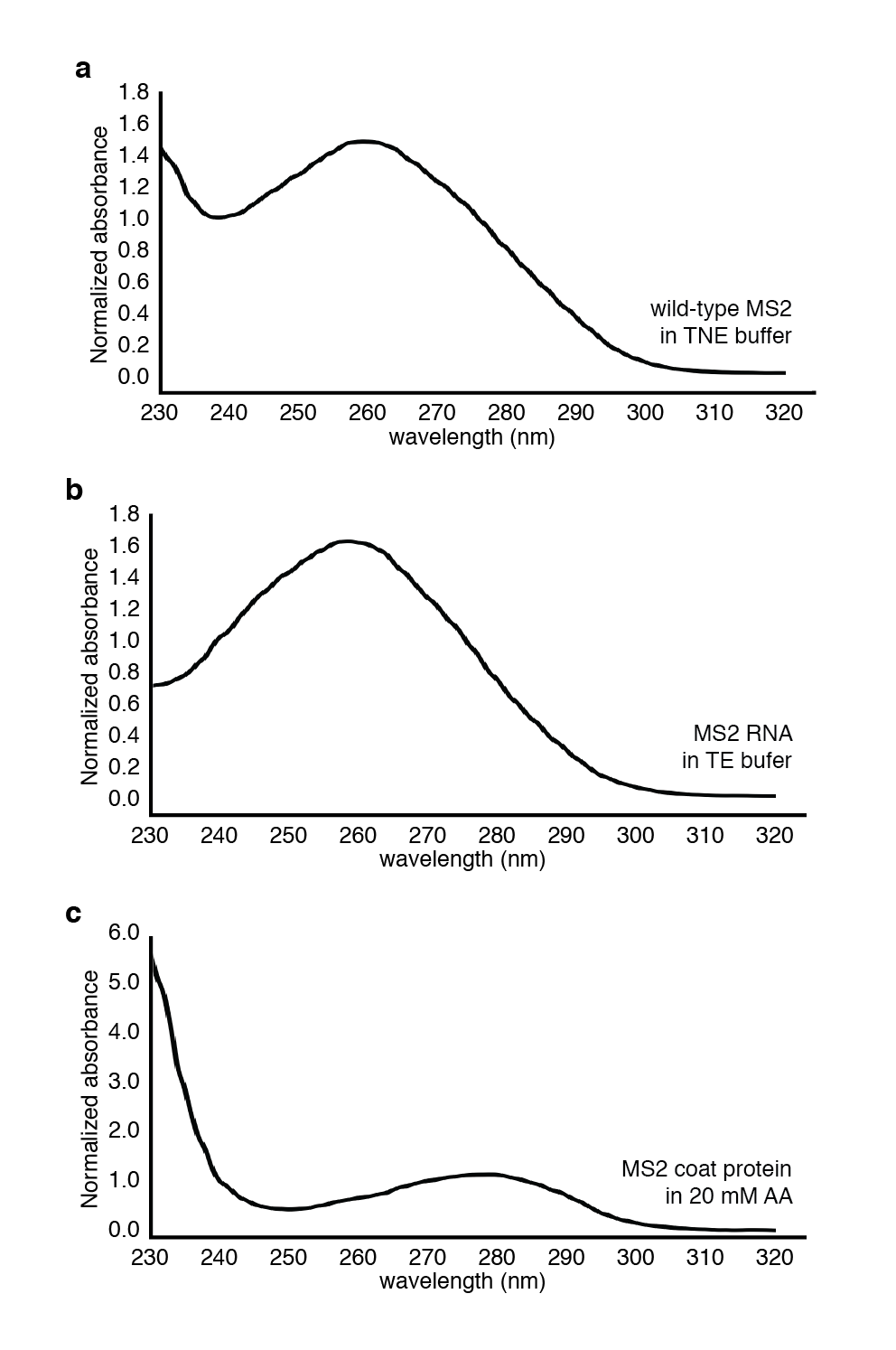}
	\caption{\textbf{Purities of wild-type MS2 virus, its RNA,
      and its coat protein are determined by UV-vis spectrophotometry.}
    Absorbance spectra for purified (a) wild-type MS2, (b) MS2
    RNA, and (c) MS2 coat protein. The 260/280 ratio of
    wild-type MS2 in TNE buffer is 1.84, of MS2 RNA in TE
    buffer is 2.16, and of unassembled coat-protein dimers in 20 mM
    acetic acid is 0.58. Each spectrum is normalized so that the
    absorbance is 1.0 at 240 nm. All absorbance measurements are made
    using a Nanodrop-1000 spectrophotometer (Thermo Scientific).}
	\label{fig:UV-vis}
\end{figure}

After purifying the coat protein from wild-type virus particles using
cold acetic acid, we exchange the coat-protein buffer for 20 mM acetic
acid using 3-kDa-MWCO centrifugal filter units (EMD Millipore). In 20~mM
acetic acid, the coat proteins form non-covalent dimers. We determine
the concentration of coat-protein dimers by UV-spectrophotometry, using
an extinction coefficient of 33200 M$^{-1}$~cm$^{-1}$ at 280 nm
(Ref.~\citenum{borodavka_evidence_2012}). We check for RNA contamination
by measuring the ratio of the UV-absorbance at 260 nm to that at 280 nm
(Supplementary Figure~\ref{fig:UV-vis}). We use only protein that
has a 260/280 ratio less than 0.67 for assembly.

\begin{figure}
  \begin{adjustwidth}{-1in}{-1in}
	\includegraphics{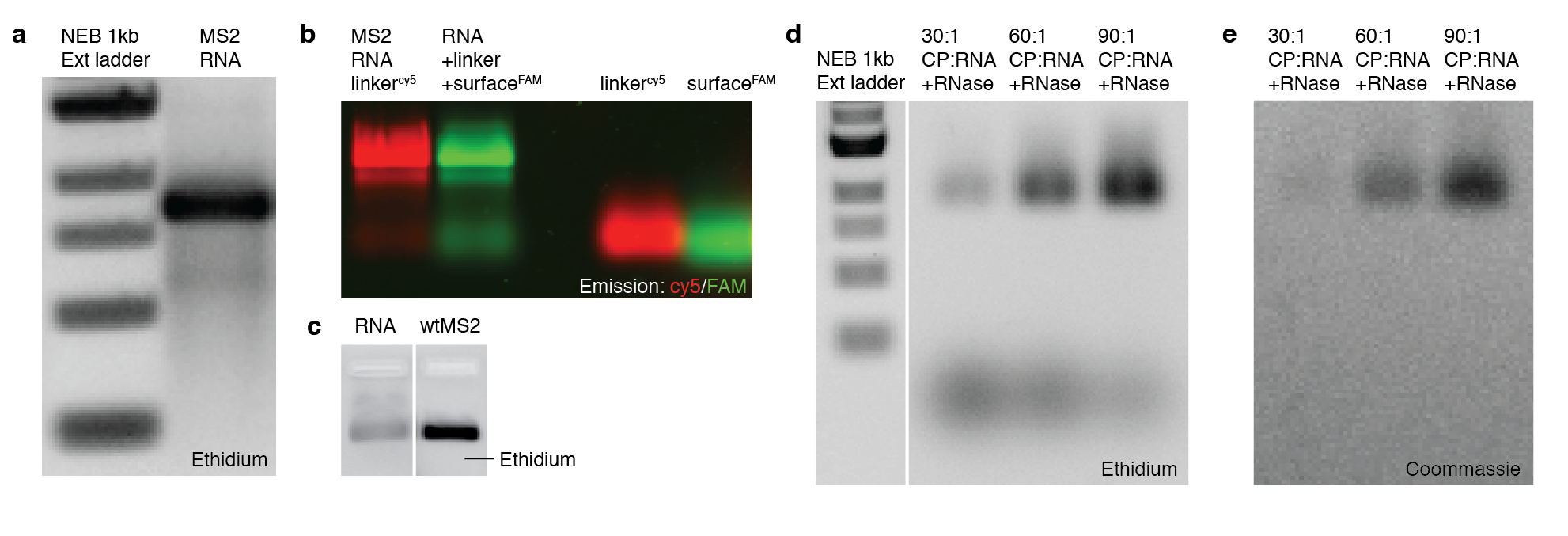}
	\caption{\textbf{Native agarose gel electrophoresis is used to
      determine the integrity of the RNA, the yield of RNA-DNA
      hybridization, and the yield of RNA packaging by MS2 coat
      protein.} All gels consist of 1\% agarose in TAE buffer. (a) The
    MS2 RNA used in the assembly experiments appears as a single band
    with minimal smearing, indicating that the RNA is not degraded. The
    left lane contains a 1-kb extended DNA ladder (New England Biolabs),
    and the right lane contains 1 $\upmu$g of MS2 RNA. The gel is
    visualized after staining with Gel Red (Biotium Inc.) ethidium
    stain. (b) Fluorescent linker and surface oligos migrate with the
    RNA after hybridization and purification, indicating strong specific
    binding. The leftmost lane is prepared by mixing 1 $\upmu$g of MS2
    RNA and a 10-fold molar excess of fluorescently labeled (5'-cy5)
    linker oligo (Integrated DNA Technologies). The RNA is hybridized to
    the linker by thermal annealing, and the unbound linker is removed
    by centrifugal filtration. The second-to-leftmost lane is prepared
    by mixing 1 $\upmu$g of MS2 RNA and a 10-fold molar excess of
    non-fluorescent linker oligo. The RNA and linker oligo are
    hybridized and the unbound linker purified as before. Then a
    stoichiometric amount of fluorescently labeled (5'-FAM) surface
    oligo (Integrated DNA Technologies) is added. The
    second-to-rightmost lane contains free 5'-cy5 linker oligo, and the
    rightmost contains free 5'-FAM surface oligo. The gel is visualized
    without staining by imaging the fluorescence emission of the cy5 and
    FAM dyes on separate channels. (c) MS2 RNA and wild-type virus
    particles migrate to the same position in the gel. The left lane
    contains RNA, and the right lane contains virus particles. The gel
    is visualized after staining with ethidium. (d) MS2 coat-protein
    dimers (CP) package MS2 RNA into RNase protected complexes with the
    same mobility as wild-type virus particles. The leftmost lane
    contains 1-kb extended ladder. The next three lanes are prepared by
    mixing 1 $\upmu$g of MS2 RNA and increasing molar ratios of CP in 10
    $\upmu$L of TNE buffer. The mixtures are incubated for 30 min at
    room temperature and then treated with 10 ng of RNase A (Amresco
    Inc.). Electrophoresis is performed 30 min after RNase treatment,
    and the gel is visualized after staining with ethidium. Protected
    RNA migrates with the same mobility as wild-type virus particles,
    and digested RNA migrates farther down the gel. The amount of
    digested RNA decreases with increasing CP. (e) Assembling particles
    prepared and then treated with RNase as just described contain
    protein, as evidenced by staining with coomassie (Instant Blue)
    protein stain.}
  	\label{fig:gels}
  \end{adjustwidth}
\end{figure}

After purifying the RNA from wild-type virus particles using an
RNeasy kit, we collect the RNA in TE buffer (10 mM Tris-HCl, pH 7.5; 1
mM EDTA) and determine its concentration by UV-spectrophotometry using
an extinction coefficient of 0.025 mL~mg$^{-1}$~cm$^{-1}$ at 260~nm. We
check for protein contamination by measuring the ratio of the
UV-absorbance at 260 nm to that at 280 nm (Supplementary
Figure~\ref{fig:UV-vis}). We use only RNA that has a 260/280 ratio
greater than 2.0 for assembly. Then we check the integrity of the RNA by
native 1\% agarose gel electrophoresis (Supplementary
Figure~\ref{fig:gels}).

\section*{Criteria for rejecting spots for analysis} 

The spots from particles that adsorb to the coverslip are easily
identified because they appear instantaneously in one frame of
the movie instead of gradually appearing over the course of many frames.
In some cases, such particles can be seen approaching the coverslip
before adsorption.

Spots within 8 pixels of the gold particle used for active stabilization
or a bright defect on the coverslip are rejected. There are typically
fewer than 2 defects on the coverslip in a given field of view. The
spots that grow near the gold particle or defect are not analyzed
because they may be due to growth that occurs on the gold particle or
defect instead of on the RNA. Furthermore, the in-plane active
stabilization keeps the coverslip position constant to within only a few
nanometers, and when particles as bright as the gold particles move by a
few nanometers they produce intensity changes that are similar to or
larger than the intensity of an MS2 capsid. These intensity changes
affect the measured intensity of any nearby assembling particles.

To determine if a spot is near a particle that adsorbs to or desorbs
from the coverslip, we check if the interference fringes of an absorbing
or desorbing particle overlap with the spot at any point during the
movie. If they do, we examine the intensity of the particle as a
function of time to check if there is an abrupt change in intensity that
occurs on the same frame as the adsorption or desorption event. If the
abrupt change in intensity is greater than 0.0003 (10\% of the intensity
of a capsid), we reject the spot for analysis. By not analyzing these
spots, we avoid misinterpreting intensity changes that are due to the
adsorption or desorption event as features of the assembly kinetics.

A spot is determined to be too close to another spot if their centers
are within 4 pixels of each other. If two spots are closer than this
distance, their interference fringes overlap, and the measured intensity
of each will depend on the intensity of the other.

Similarly, we do not analyze any spot with a center that is within 4
pixels of the edge of the field of view. We do not analyze these spots
because the interference patterns for the spot are not fully visible,
and we cannot determine if there are particles beyond the edge of the
field of view that affect the spot's intensity.

\begin{figure}
  \begin{adjustwidth}{-1in}{-1in}
	\centering
  \includegraphics{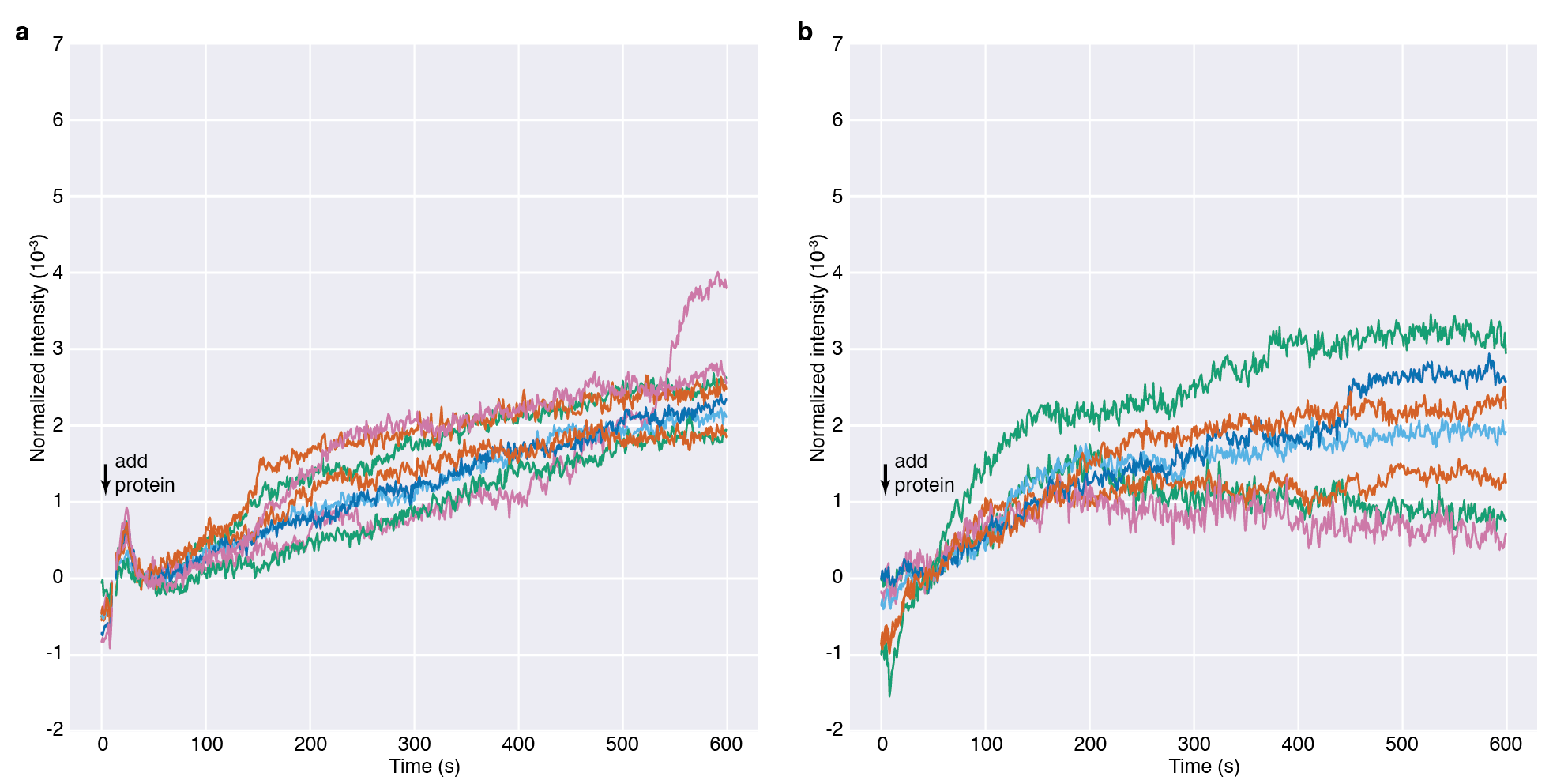}
  \caption{\textbf{Some of the spots that appear in assembly experiments
      do not represent assembly around RNA.} (a) In the assembly
    experiment from the main text where we use 2~$\upmu$M coat-protein
    dimers, we observe 8 spots (in addition to the 56 assembling
    particles described in the text) that grow slowly and synchronously
    and that show a consistent growth rate over the course of the
    measurement. The traces of these particles are shown in the plot,
    which is a 1,000-frame average of the intensities measured from
    Supplementary Movie~1, recorded at 1,000 Hz. (b) In a control
    experiment with 2~$\upmu$M dimers but no RNA on the surface, we
    observe 7 spots that grow slowly and synchronously, with traces
    similar to those shown in panel (a). For this experiment, we bound
    the linker oligos to the surface oligos, but we did not add the RNA.
    The traces are measured from the data in Supplementary Movie~5. The
    data is recorded at 1,000 Hz and is plotted with a 1,000-frame
    average.}
	\label{fig:allslowgrows}
  \end{adjustwidth}
\end{figure}

Finally, we do not analyze spots that grow slowly and synchronously with
a consistent growth rate over the course of the measurement
(Supplementary Figure~\ref{fig:allslowgrows}). In a typical experiment,
we observe 1--10 of these spots (Supplementary Movie~1). We observe a
similar number of spots with similar growth kinetics in control
experiments where RNA is not added to the surface (Supplementary
Movie~5). We therefore conclude that these spots likely do \emph{not}
represent the assembly of coat-protein dimers around RNA. They may
represent protein aggregates growing on the coverslip surface.
    
\section*{Control assembly experiments at lower
  illumination intensity} 

The results of control experiments at lower illumination intensity and 2
$\upmu$M protein shown in Supplementary
Figure~\ref{fig:control_intensity} are similar to those of the
higher-intensity experiment presented in Figure~2 of the main text and
Extended Data Figures~2 and 5. Again, different assembling particles
appear after different start times. The cumulative distribution function
of the start times is well-fit by the same exponential function but with
$t_0 =$ 62 $\pm$ 1~s, $A =$ $39.08~\substack{+0.04 \\ -0.03}$, and $\tau
=$ 49 $\pm$ 1~s for the first control experiment of the duplicate set,
and $t_0 =$ 148 $\pm$ 2~s, $A =$ 38.5 $\pm$ 0.2, and $\tau =$ 159 $\pm$
4~s for the second control experiment. 24 out of 39 traces plateau at
intensities consistent with that of a full capsid, 2 plateau at smaller
intensities, and 13 plateau at larger Intensities in the first control
experiment, while 25 out of 36 traces plateau at intensities consistent
with that of a full capsid, 5 plateau at smaller intensities, and 6
plateau at larger intensities in the second experiment. These fractions
are similar to those observed in the 2 $\upmu$M experiment presented in
the main text.

The results of the control experiments indicate that the incident light
does not qualitatively affect the assembly process. The observed kinetic
traces and distribution of start times are consistent with those
expected from a nucleation-and-growth process. Moreover, because the
difference between identically performed low-intensity control
experiments is larger than those between the high-intensity experiments
and either of the controls, we conclude that other factors, such as
differences in the concentration of protein, are responsible for the
variation. Indeed, the variation in both the fitted time constants,
$\tau$, and the delay times, $t_0$ among different experiments is not
unexpected, given the strong dependence of the start times on
concentration. At 1 $\upmu$M protein concentration, all the start times
are longer than the 600-s duration of the experiment, so that even a
slight difference in the protein concentration introduced during the 2
$\upmu$M experiments could cause the 110-s spread between the measured
time constants and the 86-s spread in the delay times.

\begin{figure}
  \begin{adjustwidth}{-1in}{-1in}
    \centering
    \includegraphics{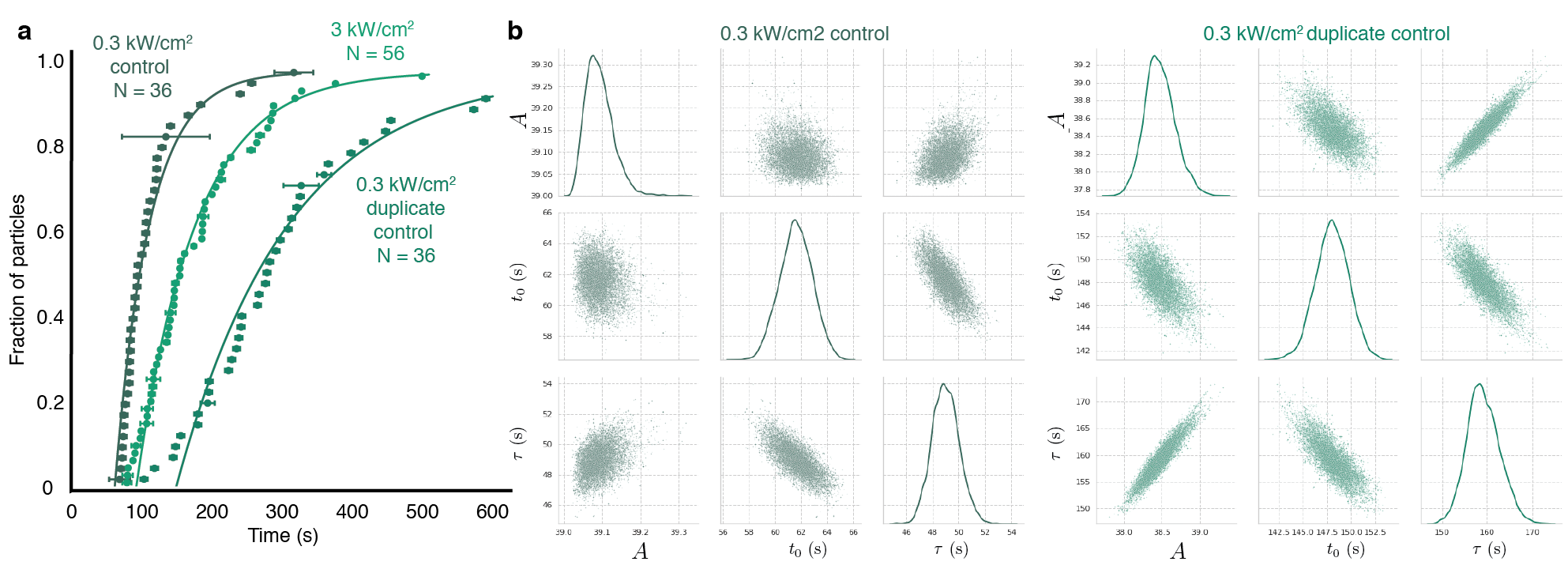}
    \caption{\textbf{Comparison of the nucleation kinetics for
      independent assembly experiments performed with 2 $\upmu$M
      of coat-protein dimers}. (a) To test if the intensity of the
    illumination beam affects the assembly process, we compare the
    cumulative distribution of start times for the experiment described
    in Figure~2 to that
    of a set of duplicate control experiments, where
    the illumination intensity is an order of magnitude smaller. The
    difference in the characteristic times for the duplicate control
    experiments is larger than the difference between the higher
    illumination intensity experiments and either of the controls,
    suggesting that other experimental uncertainties, such as
    differences in the injected protein concentration or in the flow
    profile within the imaging chamber, have a larger affect on the
    kinetics than the illumination intensity. The error bars represent
    the uncertainty in the time measurement, as described in the
    Methods. (b) The posterior probability distributions of
    parameter values obtained by fitting the data from the control
    experiments. The plots along the diagonal show kernel density
    estimates of the fully marginalized posterior distributions of each
    parameter, while the off-diagonal plots show the joint
    distributions.}
	\label{fig:control_intensity}
  \end{adjustwidth}
\end{figure}

\clearpage

\section*{Supplementary Movies}

{\raggedleft{}\textbf{Supplementary Movie 1}: The time-series of images
  from the assembly experiment using 2~$\upmu$M protein (Figs.~1,~2,
  and~3, Extended Data Fig.~2, and Supplementary Figs. 6 and 7). The
  time-series is recorded at 1,000 Hz, is shown with a 1,000-frame
  average, and is sped up by a factor of 100 for playback. The field of
  view is 9.8 $\upmu$m on each side.}\\
  
{\raggedleft{}\textbf{Supplementary Movie 2}: The time-series of images
  from the assembly experiment using 1 $\upmu$M protein (Extended
  Data Fig.~6). The time-series is recorded at 100 Hz, is shown with a
  300-frame average, and is sped up by a factor of 200 for playback. The
  field of view is 14 $\upmu$m on each side. The illumination beam is
  blocked for a short time approximately halfway through the movie, just
  before 2 $\upmu$M protein is added. In the first half of the movie,
  where 1 $\upmu$M protein is in the imaging chamber, a few particles
  are seen adsorbing to the coverslip, but no particles are seen growing
  on the coverslip. In the second half of the movie, where 2 $\upmu$M of
  protein is in the imaging chamber, a number of particles are seen
  growing on the
  coverslip.}\\
  
{\raggedleft{}\textbf{Supplementary Movie 3}: The time-series of images
  from the assembly experiment using 1.5~$\upmu$M protein (Fig.~3,
  Extended Data Fig.~7). The time-series is recorded at 1,000 Hz, is
  shown with a 1,000-frame average, and is sped up by a factor of 100
  for playback. The field of
  view is 9.8 $\upmu$m on each side.}\\

{\raggedleft{}\textbf{Supplementary Movie 4}: The time-series of images
  from the assembly experiment using 4~$\upmu$M protein (Fig.~3,
  Extended Data Fig.~8). The time-series is recorded at 1,000 Hz, is
  shown with a 1,000-frame average, and is sped up by a factor of 100
  for playback. The field of
  view is 9.8 $\upmu$m on each side.}\\
  
  {\raggedleft{}\textbf{Supplementary Movie 5}: The time-series of images
  from the control experiment using 2~$\upmu$M protein with no RNA on
  the coverslip (Supplementary Fig. 6). The
  time-series is recorded at 1,000 Hz, is shown with a 1,000-frame
  average, and is sped up by a factor of 100 for playback. The field of
  view is 9.8 $\upmu$m on each side.}\\ 

\bibliographystyle{achemso}
\bibliography{virus-assembly}